\documentclass[11pt]{article}
\usepackage{amssymb,amsfonts,amsmath,amsthm}
\usepackage{graphicx}
\usepackage{setspace}
\usepackage{epstopdf}
\usepackage{pdflscape}
\usepackage{cite}
\usepackage{natbib}
\usepackage[capposition=top]{floatrow}
\usepackage[inline]{enumitem}
\usepackage{bm}

\newtheorem{thm}{Theorem}
\newtheorem{coro}[thm]{Corollary}

\newtheorem{lemma}[thm]{Lemma}
\newtheorem{assump}{Assumption}

\DeclareMathOperator*{\sgn}{sgn}

\makeatletter
\renewenvironment{proof}[1][\proofname]{%
  \par\pushQED{\qed}\normalfont%
  \topsep6\p@\@plus6\p@\relax
  \trivlist\item[\hskip\labelsep\bfseries#1\@addpunct{.}]%
  \ignorespaces
}{%
  \popQED\endtrivlist\@endpefalse
}
\makeatother

\providecommand{\abs}[1]{\left\lvert#1\right\rvert}
\providecommand{\norm}[1]{\left\lVert#1\right\rVert}

\title{Analysis of Distributional Dynamics for Repeated Cross-Sectional and Intra-Period Observations\thanks{We are grateful for useful comments to Yoosoon Chang, Chang Sik Kim, Peter Phillips, Marine Carrasco and the participants of 2015 Midwest Econometrics Group Meeting, 2016 SETA, Time Series Workshop on Macro and Financial Economics, International Symposium on Financial Engineering and Risk Management, SoFiE, CMES, and the attendants of the seminars at Toulouse School of Economics, Einaudi Institute of Economics and Finance, University of Cincinnati, University of Montreal and Indiana University. Junhui Qian would like to thank National Natural Science Foundation of China for financial support (Project ID: 71673183).}}
\author{Bo Hu\smallskip \\ {\small Institute of New Structural Economics}\\{\small Peking University}
\and Joon Y. Park\smallskip \\{\small Department of Economics}\\{\small Indiana University}\\{\small and\! Sungkyunkwan\! University}\medskip\\
\and Junhui Qian\smallskip \\ {\small School of Economics}\\
{\small Shanghai\! Jiao\! Tong\! University}}
\date{February 20, 2020}

\begin{document}

\maketitle

\begin{abstract}
This paper introduces a novel approach to investigate the dynamics of state distributions, which accommodate both cross-sectional distributions of repeated panels and intra-period distributions of a time series observed at high frequency. In our approach, densities of the state distributions are regarded as functional elements in a Hilbert space, and are assumed to follow a functional autoregressive model. We propose an estimator for the autoregressive operator, establish its consistency, and provide tools and asymptotics to analyze the forecast of state density and the moment dynamics of state distributions. We apply our methodology to study the time series of distributions of the GBP/USD exchange rate intra-month returns and the time series of cross-sectional distributions of the NYSE stocks monthly returns. Finally, we conduct simulations to evaluate the density forecasts based on our model.
\end{abstract}
\vfill

\noindent JEL classification codes: C13, C14, C22 \smallskip \\
\noindent Key words and phrases: distributional dynamics, repeated panels, intra-period distributions, functional autoregression, time-varying density, density forecasting

\newpage
\onehalfspacing
\section{Introduction}
Many important issues in economics are related to the time evolution of state distributions of economic variables, which are defined to be either cross-sectional or intra-period distributions of the underlying economic variables. For example, the time series of cross-sectional or intra-period distributions of returns from traded assets contain important information on the dynamics of risk and return in financial markets. However, tracking the complete distributional dynamics is not an easy task because of the infinite dimensional nature of state distributions. The conventional approaches usually get around the problem by only considering some particular dimensions of the distributions such as the mean, the variance, or some specific quantiles. Information beyond these dimensions, however, would be necessarily ignored. In this paper, we develop a novel analytical framework that can be used to track and exploit the dynamics of entire state distributions.

We consider the functional autoregressive (FAR) model of time-varying distributions. Allowing the distributions to be infinite-dimensional, we adopt a nonparametric approach to modeling distributional dynamics. The parametric approach has been well studied, including the autoregressive conditional moment models by \citet{engle-82}, \citet{bollerslev-86},  \citet{engle-lilien-robins-87}, \citet{harvey-siddique-99} and \citet{brooks-burke-heravi-persand-05}, the autoregressive conditional density model by \citet{hansen-94}, and the autoregressive quantile models by \citet{koenker-xiao-06} and \citet{xiao-koenker-09}, among many others. Our approach, on the contrary, does not require any parametric assumptions on the underlying state distributions. Nevertheless, it allows us to investigate full temporal dynamics of state distributions, including dependence structures among moments and tail probabilities, within a single framework.

For our analysis, we represent state distributions by their density functions and regard them as random elements of a Hilbert space. Of course, it is also possible to use other functions such as distribution functions and quantile functions to represent state distributions.\footnote{\citet{boogaart-egozcue-pawlowsky-glahn-14} propose to use the so-called Bayes Hilbert space, which uses log-densities instead of densities themselves, to study distributional dynamics. Though their approach has some advantages, it does not allow for any of the methodologies developed in our paper to analyze distributional dynamics.} However, we choose density functions, since they provide clear and direct interpretations on the dynamics of moments, tail probabilities and other important features of state distributions. Indeed, in our framework, the cross-sectional moments and tail probabilities are defined simply as the inner products of state density functions with some deterministic functions in a Hilbert space. No other functions representing state distributions have this nice property. We fit our FAR model using the time series of state density functions, which are obtained from repeated cross-sectional and intra-period high frequency observations.\footnote{In all our applications, the number of cross-sectional or intra-period observations is relatively much bigger than the number of time series observations. Therefore, we ignore the statistical errors incurred by the estimation of state densities using repeated cross-sectional and intra-period observations.} Once fitted, we estimate the leading progressive/regressive features and conduct various impulse response analyses to reveal the way in which future state distributions respond to changes in past state distributions. In addition, we develop a variance decomposition method that relates a specific feature of a current state distribution to the moments of its past. Lastly, we propose a density predictor and find its asymptotic properties.

Recently, there has been a rapid development in the theory and applications of functional time series analysis. Many authors, including \citet{besse-cardot-stephenson-00}, \citet{bosq-00}, \citet{antoniaids-paparoditis-sapatinas-06}, \citet{ferraty-vieu-06}, \citet{mas-07}, \citet{kargin-onatski-08}, \citet{hormann-kokoszka-10}, \citet{horvath-huskova-kokoszka-10}, \citet{didericksen-kokoszka-zhang-12}, \citet{kokoszka-reimherr-13}, \citet{panaretos-tavakoli-13a}, \citet{panaretos-tavakoli-13b}, \citet{aue-norinho-hormann-15}, \citet{aue-horvath-pellatt-17} and \citet{klepsch-kluppelberg-17}, among others, studied various aspects of functional data. In many of these works, however, temporal dependencies in the functional dynamics are used in a relatively implicit way to make forecasts. We in this paper emphasize the investigation of the functional dynamics themselves in the distributional dynamics setting, and develop tools to interpret them in economically meaningful manners.

We apply our methodology to two representative state distributions: the \emph{intra-month} distributions of the UK Pound/US Dollar (GBP/USD) exchange rate 15-minute log returns and the \emph{cross-sectional} distributions of the New York Stock Exchange (NYSE) stocks monthly returns. In both cases, we find that observations/stocks with small returns play the most important role in interacting with the past distribution and in determining the future distribution. Outliers turn out to be unimportant in the distributional dynamics. We also find that various characteristics such as the moments and the tail probabilities of the current distribution respond differently to shocks to, and subsequent changes in, the relative frequency of observations with different levels of returns. The momentum effect appears in the dynamics of the first moment: The more observations with positive (negative) returns in the past distribution, the higher (lower) the first moment of the current distribution. On the other hand, sizes of the returns play a more important role in the dynamics of the second moment; the more observations with small (large) returns in the past distribution, the lower (higher) the second moment of the current distribution.

In contrast to the symmetry found in the behaviors of left and right tail probabilities in the foreign exchange rate return application, those in the stock return application appear to be quite asymmetric. In the former, a higher relative frequency of large returns in the past distribution, regardless of the signs of these returns, increases both the current left and right tail probabilities rather symmetrically. In the latter, however, the current left tail probability is mostly determined by the relative frequency of moderate-to-large negative returns in the past distribution, whereas the current right tail probability is positively affected both by the relative frequency of moderate-to-large negative returns, though to a lesser degree, and by the relative frequency of positive returns in the past distribution. The variance decomposition exercises indicate that in both cases, the variances of the current first moments are not explained much by the variances of the past integral moments, while the current second moments are more closely connected to the past second moments. These results lend support to the usual observation that in financial markets, the mean is not quite predictable while the volatility is much more persistent, and thus predictable. These moment dependence findings are also consistent with the rolling out-of-sample forecast results using our model.

The rest of the paper is organized as follows. In Section 2, we introduce the functional autoregressive model of time-varying densities and the necessary theoretical background for understanding the model. Section 3 explains how we may estimate the model and  develops the relevant statistical theory. In particular, our estimators of the autoregressive operator and of the error variance operator are shown to be consistent under appropriate regularity conditions. Section 4 presents density forecast along with its asymptotic theory, and provides a variance decomposition method that is useful in studying the moment dynamics of state distributions. In Section 5, we apply our model and methodology to the state distributions representing intra-month distributions of high frequency GBP/USD exchange rate returns and cross-sectional distributions of the NYSE stocks monthly returns. Section 6 summarizes our simulation results on the finite sample performance of the density predictor. Section 7 concludes the paper. All mathematical proofs are collected in Appendix.

\section{The Model and Preliminaries}
In this section, we present our model and some preliminaries that are necessary for the development of our theory and methodology in the paper.

\subsection{The Model}
In this paper, we consider a random sequence $(f_t)$ of probability densities of state distributions, which represent some distributional aspect of an economy, for time $t=1,2,\ldots$. The density functions $(f_t)$ are time varying, and are regarded as random elements taking values in the Hilbert space $L^2(C)$ of square integrable functions on some compact subset $C$ of $\mathbb{R}$ or $\mathbb R$ itself. We may introduce the underlying probability space $(\Omega, \mathcal{F}, \mathbb{P})$, and more formally define $f_t:\Omega\to L^2(C)$ for $t=1,2,\ldots$ with the required measurability condition.

We assume that $(f_t)$ has a well defined common mean $\mathbb{E} f$, whose precise meaning will be introduced later in Section \ref{sec:prelim}. Moreover, we let the demeaned state densities $(w_t)$, $w_t= f_t - \mathbb{E} f$, be generated according to
\begin{equation}
  \label{eq:model}
  w_t = Aw_{t-1} + \varepsilon_t
\end{equation}
for $t = 1, 2, \ldots$, where $A$ is a bounded linear operator, called the autoregressive operator, and $(\varepsilon_t)$ is a functional white noise process.\footnote{Of course, we may consider the model $f_t = \mu + Af_{t-1} + \varepsilon_t$ given directly in terms of the state densities $(f_t)$ themselves instead of the demeaned state densities $(w_t)$, where $\mu$ is an additional unknown parameter function corresponding to a constant term in the usual regression. In this paper, we use a more compact presentation in \eqref{eq:model} to focus on the estimation and interpretation of the operator $A$.} The meaning of a functional white noise process will be introduced later in Section \ref{sec:prelim}. Note that $(w_t)$ takes values in a subset $H$ of $L^2(C)$, which is given by\footnote{Here and elsewhere in this paper, we use the notation $1$ to denote the constant function taking value $1$, instead of the identity function.}
\[
 H = \left\{v\in L^2(C)\big|\langle 1, v\rangle = 0\right\},
\]
where the inner product in the Hilbert space $H$ is defined by $\langle z, w\rangle = \int_C z(x)w(x)dx$, and $1$ denotes the constant function taking value $1$ everywhere on the support. Throughout the paper, we assume that $A$ is an operator on $H$, so that $(\varepsilon_t)$ also take values in $H$.

The model introduced in \eqref{eq:model} may be regarded as a special example of functional autoregression (of order 1), which is often denoted by FAR(1). From this view, our time series $(w_t)$ of demeaned state densities is just a special functional autoregressive process (of order 1). We may define more general FAR($p$) models and processes in a similar way. FAR has the same structure as, and may thus be regarded as, a generalization of the vector autoregression (VAR) that has been widely used in time series econometric modeling. In our FAR model, we simply have functional variables, which are generally infinite dimensional. This contrasts to VAR, where we have a finite number of variables included in the model. FAR and VAR are the same in many aspects such as motivation and intended interpretation. However, technically it is more difficult and involved to deal with FAR than with VAR. For instance, the infinite dimensionality of FAR introduces the ill-posed inverse problem, which makes it harder to do inference. Interested readers are referred to \cite{bosq-00} and the references therein for more detailed exposition on FAR.

One advantage of our FAR model for state densities introduced in \eqref{eq:model} is that it may be used to investigate the intertemporal dynamics of the moments of state distributions represented by their densities. To explain our approach, let $v\in H$, and consider a coordinate-process version of our model given as
\begin{equation}
  \label{eq:model_coor}
  \langle v, w_t\rangle = \langle v, Aw_{t-1}\rangle + \langle v, \varepsilon_t \rangle = \langle A^*v, w_{t-1}\rangle + \varepsilon_t(v),
\end{equation}
where $A^*$ is the adjoint of $A$ and $\big(\varepsilon_t(v)\big)$ is a scalar white noise process. The joint operator of a linear operator $A$ on a Hilbert space $H$ is the linear operator $A^*$ on $H$ such that for any $z, w\in H$, $\langle z, Aw\rangle = \langle A^*z, w\rangle$. For any $v\in H$ given, we may interpret $A^\ast v$ as the response function of $\langle v,w_t\rangle$ to an impulse to $w_{t-1}$ given by a Dirac delta function. Note that $\langle v,w_t\rangle$ would increase by $\langle A^\ast v,\delta_x\rangle = (A^\ast v)(x)$, in response to an impulse to $w_{t-1}$ given by the Dirac delta function $\delta_x$ with a spike at $x$.

We may use \eqref{eq:model_coor} to analyze the intertemporal dynamics of various moments of state distributions.\footnote{Here by convention we extend the inner product in $H$ to define $\langle v,w_t\rangle$ for $v\in L^2(C)$. This convention will be made throughout the paper.} For $\iota_p$ defined as $\iota_p(x) = x^p$ for $p=1,2,\ldots$, we have
\begin{equation*}
 \langle\iota_p,w_t\rangle = \int_C x^p w_t(x)dx,
\end{equation*}
which is the $p$-th moment of the demeaned state distribution at time $t$. Furthermore, if we write $A^*\iota_p = \sum_{q=1}^\infty c_{p,q}\iota_q$ for a given $p$ with some real sequence $(c_{p,q})$ for $q=1,2,\ldots$, then it follows that
\begin{equation*}
 \langle A^\ast\iota_p,w_{t-1}\rangle
 = \sum_{q=1}^\infty c_{p,q} \int_Cx^qw_{t-1}(x)dx,
\end{equation*}
which is an infinite linear combination of all integral moments of the lagged demeaned state distribution represented by $(w_{t-1})$. Consequently, it follows from \eqref{eq:model_coor} that
\[
 \int_C x^p w_t(x)dx = \sum_{q=1}^\infty c_{p,q} \int_Cx^qw_{t-1}(x)dx + \varepsilon_t(\iota_p),
\]
which can be used to analyze the moment dynamics of state distributions.

Two special cases appear to be worth mentioning. In what follows, we let $(\mathcal F_t)$ be a filtration, to which $(w_t)$ is adapted. If we assume $A^\ast\iota_2 = \alpha\iota_2$ for some constant $0<\alpha<1$, it follows that $\langle\iota_2, w_t\rangle = \alpha\langle\iota_2, w_{t-1}\rangle + \varepsilon_t(\iota_2)$, from which we may deduce that
\[
 \mathbb E\left(\langle\iota_2, w_t\rangle\big|\mathcal F_{t-1}\right)
 = \alpha\langle\iota_2, w_{t-1}\rangle.
\]
Therefore, \eqref{eq:model_coor} with $v=\iota_2$ reduces essentially to an ARCH model. If in addition to $A^\ast\iota_2 = \alpha\iota_2$, we let $A^\ast\iota_1 = \alpha_1\iota_1 + \alpha_2\iota_2$, then we have
\[
 \langle\iota_1,w_t\rangle = \beta_1\langle\iota_1,w_{t-1}\rangle
 + \beta_2\mathbb E\left(\langle\iota_2,w_t\rangle\big|\mathcal F_{t-1}\right) + \varepsilon_t(\iota_1)
\]
where $\beta_1 = \alpha_1$ and $\beta_2 = \alpha_2/\alpha$. In this case, \eqref{eq:model_coor} with $v=\iota_1$ becomes an ARCH-M model studied in \cite{engle-lilien-robins-87}. Clearly, our model provides a much more general framework within which we analyze various moment dynamics of general state distributions.

In addition, we may have
\[v(x) = 1_B(x),\]
where $1_B$ is an indicator function for a subset $B$ of $C$. In this case, $\langle v, w_t\rangle = \int_B w_t(x)dx$ in \eqref{eq:model_coor} represents the probability of the underlying state being at $B$, in terms of the deviation from its expected value. More specifically, for instance, the probability of the underlying state being at a left tail can be analyzed if we set $B=(-\infty,-\tau)\cap C$ with $\tau>0$. We can also choose $B=\big((-\infty,-\tau)\cup (\tau,\infty)\big)\cap C$, so that the probability of the underlying state being at an extreme value can be studied.

Throughout this paper we make the following assumptions:

\begin{assump}
  \label{assump_1}
  We assume that
  \begin{enumerate}[label=(\alph*), nosep]
  \item $A$ is a compact linear operator such that $\|A^k\|<1$ for some $k\geq 1$, and,
  \item $(\varepsilon_t)$ are i.i.d. such that $\mathbb{E} \varepsilon_t = 0$ and $\mathbb{E}\norm{\varepsilon_t}^4<\infty$, and are independent of $w_0$.
  \end{enumerate}
\end{assump}

We use $\norm{\cdot}$ to denote the operator norm for bounded operators on $H$. The operator norm of $A$ on $H$ is defined as $\norm{A} = \sup_{v\in H} \norm{Av}/\norm{v}$. Recall that a linear operator is compact if it maps the open unit ball to a set that has a compact closure. Since a compact operator on a Hilbert space can be represented as the limit (in operator norm) of a sequence of finite dimensional operators, many features of the finite dimensional matrix theory naturally generalize to compact linear operators. As a consequence, we may regard $A$ essentially as an infinite dimensional matrix. As shown in \cite{bosq-00}, $\|A^k\|<1$ for some $k\ge 1$ if and only if $\|A^k\|\leq ab^k$ with some $a>0$ and $0<b<1$ for all $k\ge 0$.  This condition implies that FAR($1$) introduced in \eqref{eq:model} has a unique stationary solution for $(w_t)$. The i.i.d. assumption for $(\varepsilon_t)$ introduced above is more restrictive than is necessary and can be relaxed to less stringent conditions. Many of our subsequent results hold for sequences that are only serially uncorrelated.

\subsection{Some Preliminaries}\label{sec:prelim}
For an $H$-valued random variable $w$, we define its mean by the element $\mathbb{E} w\in H$ such that we have, for all $v\in H$,
\begin{equation}
  \label{eq:def_bbE}
  \langle v, \mathbb{E} w\rangle = \mathbb{E}\langle v,w\rangle.
\end{equation}
If $\mathbb{E}\|w\|<\infty$, then $\mathbb{E} w$ exists and is unique. It can be shown that many of the properties of the usual expectation hold in the functional random variable case. For example, the expectation operator is linear. In particular, if $A$ is a bounded linear operator in $H$, we have that $\mathbb{E}(Aw) = A\mathbb{E} w$. In addition, we have that $\norm{\mathbb{E} w} \leq \mathbb{E} \norm{w}$. We define the covariance operator of two $H$-valued random variables $z$ and $w$ by the linear operator $\mathbb{E}(z\otimes w)$ on $H\times H$, for which we have
\begin{equation*}
 \big[\mathbb{E}(z\otimes w)\big]v = \mathbb{E}\langle w,v\rangle z
\end{equation*}
for all $v\in H$. Note that we have utilized the tensor product notation defined by $(u\otimes v)(\cdot) = \langle v,\cdot\rangle u$ for any $u, v\in H$. If $\mathbb{E}\norm{z}^2<\infty$ and $\mathbb{E}\norm{w}^2<\infty$, then $\mathbb{E}(z\otimes w)$ exists and is unique. Naturally, we may call $\mathbb{E}(w\otimes w)$ the variance operator of $w$.

An $H$-valued random variable $w$ is called Gaussian if the random variable $\langle v, w\rangle$ is Gaussian for all $v\in H$. We denote by $\mathbb{N}(\mu, \Sigma)$ a Gaussian random variable with mean $\mu$ and variance operator $\Sigma$.

An operator $A$ on a separable Hilbert space $H$ is compact if and only if it can be written as
\begin{equation}
  \label{eq:svd_of_compact_operator}
  A = \sum_{k=1}^\infty \kappa_k(u_k\otimes v_k)
\end{equation}
for some orthonormal systems $(u_k)$ and $(v_k)$ of $H$ and a sequence $(\kappa_k)$ of nonnegative numbers tending to 0. The compact linear operator is called nuclear if $\sum_{k=1}^\infty \abs{\kappa_k} < \infty$, and Hilbert-Schmidt if $\sum_{k=1}^\infty \kappa_k^2 < \infty$. It is well known that $\mathbb{E}(z\otimes w)$ is nuclear, and therefore compact. Furthermore, the variance operator $\mathbb{E}(w\otimes w)$ is self-adjoint and nonnegative (strictly positive if non-degenerate), and admits the spectral representation
\begin{equation}
  \label{eq:3}
  \mathbb{E}(w\otimes w) = \sum_{k=1}^\infty \lambda_k(v_k\otimes v_k)
\end{equation}
for some nonnegative (strictly positive if non-degenerate) sequence $(\lambda_k)$ and some orthonormal basis $(v_k)$ of $H$.

The spectral representation of the autoregressive operator as in \eqref{eq:svd_of_compact_operator} gives rise to interesting interpretations of the autoregressive operator. Note that, under \eqref{eq:svd_of_compact_operator}, our model in \eqref{eq:model} can be written as
\begin{equation*}
  w_t = \sum_{k=1}^\infty \kappa_k \langle v_k,w_{t-1}\rangle u_k + \varepsilon_t.
\end{equation*}
If we order $(\kappa_k)$ and correspondingly $(v_k)$ and $(u_k)$ such that $\kappa_1>\kappa_2>\ldots$,
then $v_1$ may be viewed as the direction in which $w_{t-1}$ mainly affects $w_t$. The corresponding feature of the distribution generated from $v_1$ will be called the \emph{leading progressive feature}. On the other hand, $u_1$ represents the direction in which $w_t$ is mainly affected by $w_{t-1}$. The corresponding feature of the distribution generated from $u_1$ will be called the \emph{leading regressive feature}. For example, if $v_1 = \iota_2$ and $u_1 = \iota_1$ so that the leading progressive feature is the second moment and the leading regressive feature is the first moment, then in this process the second moment of $w_{t-1}$ mainly affects the distribution of $w_t$ and the first moment of $w_t$ is mainly affected by the distribution of $w_{t-1}$.

A sequence $(\varepsilon_t)$ of $H$-valued random variables is called a white noise if $\mathbb{E}(\varepsilon_t\otimes \varepsilon_t)$ is the same for all $t$ and $\mathbb{E}(\varepsilon_t \otimes \varepsilon_s)=0$ for all $t\neq s$. It is easy to see that if $(\varepsilon_t)$ is a white noise, then for any $v\in H$, the scalar process $(\varepsilon_t(v))$, $\varepsilon_t(v) = \langle v, \varepsilon_t\rangle$, is a white noise. Under Assumption \ref{assump_1}, the model \eqref{eq:model} has a unique stationary solution for $(w_t)$, which is given by
\begin{equation*}
  w_t = \sum_{k=0}^\infty A^k\varepsilon_{t-k},
\end{equation*}
where $(\varepsilon_t)$ is a white noise process.
See, for instance, \cite{bosq-00} for more details. It is therefore convenient to define the autocovariance function of $(w_t)$ by
\begin{equation*}
  \Gamma(k) = \mathbb{E}(w_t\otimes w_{t-k})
\end{equation*}
for $k\in \mathbb Z$, as in the analysis of finite dimensional time series. Of particular interest are $P = \Gamma(1)$ and $Q = \Gamma(0)$. It is easy to deduce that $\Gamma(k) = A^kQ$ for $k\geq 1$ and $\Gamma(k) = \Gamma(-k)^*$ for all $k$. In particular, we have
\begin{equation}
  \label{eq:CCR}
  P = AQ,
\end{equation}
which will be used to estimate $A$.

It is well known that compact linear operators on infinite dimensional Hilbert spaces are not invertible. Therefore one may not directly use the relationship in \eqref{eq:CCR} to define the autoregressive operator $A$ as $A = PQ^{-1}$ since $Q^{-1}$ is not well defined on $H$. In fact, if the kernel of $Q$ is $\{0\}$, then $Q^{-1}$ is well defined only on $\mathcal{R}(Q) = \{u\in H|\sum_{k=1}^\infty \langle u, v_k\rangle^2/\lambda_k^2 <\infty\}$, which is a proper subspace of $H$. Consequently, we have $A=PQ^{-1}$ well defined only on the restricted domain $\mathcal{D}=\mathcal{R}(Q)$. This problem is often referred to as the ill-posed inverse problem.

The standard method to circumvent this problem in the functional data analysis literature is to restrict the definition of $A$ in a finite dimensional subspace of $H$. To explain this method, let $\lambda_1>\lambda_2>\cdots>0$ and define $V_K$ to be the subspace of $H$ spanned by the $K$ eigenvectors $v_1, \ldots, v_K$ associated with the eigenvalues $\lambda_1, \ldots, \lambda_K$ of $Q$. Let $Q_K = \Pi_KQ\Pi_K$ where $\Pi_K$ denotes the projection onto $V_K$ and define
\begin{equation}
  \label{eq:Q_K+}
  Q_K^+ = \sum_{k=1}^K \lambda_k^{-1}(v_k\otimes v_k),
\end{equation}
i.e., the inverse of $Q$ on $V_K$.\footnote{See, e.g., \cite{benatia-carrasco-florens-17} for the use of other methods of regularization.}

Now let
\begin{equation}
  \label{eq:A_K}
  A_K = PQ_K^+,
\end{equation}
which is the autoregressive operator $A$ restricted to the subspace $V_K$ of $H$. Note that $V_K$ is generated by the first $K$ principal components of $(w_t)$. Since $(\lambda_k)$ decreases to zero, we may expect that $A_K$ approximates $A$ well if the dimension of $V_K$ increases. The estimator of $A$, which will be introduced in the next section, is indeed the sample analogue estimator of $A_K$ in \eqref{eq:A_K}, and we let $K$ increase as $T$ increases.

\section{Estimation}
In most applications, $(f_t)$ are not directly observed and should therefore be estimated before we look at the FAR model specified by \eqref{eq:model}. We suppose that $N$ observations from the probability density $f_t$ are available so that we may estimate $f_t$ consistently for each $t = 1, 2, \ldots$. In what follows, we denote by $\hat f_t$ the consistent estimator of $f_t$ based on $N$ observations and define $\Delta_t = \hat f_t - f_t$ for $t = 1, \ldots, T$. As before, we let $(\lambda_k)$ be the ordered eigenvalues of $Q$.

\begin{assump}
  \label{assump_2}
  We assume that
  \begin{enumerate}[label=(\alph*), nosep]
  \item \label{ass2_a} $\lambda_1\geq \lambda_2\geq \lambda_3\geq \cdots>0$,
  \item \label{ass2_b} $\norm{f_t}, \norm{\Delta_t}\leq M$ a.s. for some constant $M>0$, and,
  \item \label{ass2_c} $\sup_{t\geq 1} \mathbb{E}\norm{\Delta_t}^2 = O(N^{-r})$ for some $r>0$.
  \end{enumerate}
\end{assump}

Assumption \ref{assump_2} is very mild and appears to be widely satisfied in practical applications. The condition in part \ref{ass2_a} holds if and only if the kernel of $Q$ contains only the origin, which is met as long as the distribution of $(f_t)$ is non-degenerate. Boundedness of $(f_t)$ in part \ref{ass2_b} is expected to hold in many practical applications, since $(f_t)$ is a family of density functions. This is not strictly necessary, and is only required for an optimal rate in the strong laws of large numbers we establish in Section 3.1 and 3.2. For unbounded $(f_t)$, we have a reduced rate of convergence for the strong laws of large numbers. However, the $L^2$-convergence results in Section 3.1 and 3.2 hold for unbounded $(f_t)$ without any modification. The condition in part (c) holds in many applications. For instance, if $(f_t)$ are estimated for each $t$ by the kernel density estimator from $N$ observations, we may expect for all $t$ that $\mathbb{E}\norm{\Delta_t}^2 = O(N^{-r})$ for some $0<r<2/5$ under very general regularity conditions which allow in particular for dependency among $N$ observations. See, e.g., \citet[Theorem 2.2]{bosq-98}. One may expect even higher rates of convergence if higher-order kernels are used. See, e.g., \citet[Section 2.8]{wand-jones-95} .

\subsection{Mean and Autocovariance Operators}
\label{sec:variance_est}
First we estimate $\mathbb{E} f$ by the sample average of $(f_t)$, i.e., by
\begin{equation*}
  \bar f = \frac{1}{T}\sum_{t=1}^T \hat f_t.
\end{equation*}
The following result holds.

\begin{lemma}
  \label{lemma:1}
  Let Assumptions 1 and 2 hold. If $N\geq cT^{1/r}$ for some constant $c>0$, then we have
  \begin{equation*}
    \mathbb{E}\norm{\bar f-\mathbb{E} f}^2 = O\left( T^{-1} \right)
  \end{equation*}
  as $T\to \infty$. Furthermore, if $N>cT^{2/r}\ln^s T$ for some constants $c>0$ and $s>0$, then we have
  \begin{equation*}
    \norm{\bar f- \mathbb{E} f} = O\left(T^{-1/2}\ln^{1/2}T \right) \quad \text{a.s.}
  \end{equation*}
as $T\to \infty$.
\end{lemma}

Lemma \ref{lemma:1} establishes the $L^2$ and a.s. consistency for the sample mean $\bar f$. It shows in particular that using estimated densities $(\hat f_t)$ in place of $(f_t)$ does not affect the convergence rates established in Theorem 3.7 and Corollary 3.2 of \cite{bosq-00} as long as the number of observations used to estimate $(f_t)$ is sufficiently large. The conditions in Assumptions \ref{assump_1} and \ref{assump_2} are sufficient to yield optimal rates in the laws of large numbers here.

Similarly, we may use the sample analogue to consistently estimate the autocovariance operators $Q$ and $P$.  Let $\hat w_t = \hat f_t - \bar f$ for $t = 1, 2, \ldots$, and define
\begin{equation*}
  \hat Q = \frac{1}{T}\sum_{t=1}^T(\hat w_t\otimes \hat w_t)
\end{equation*}
and
\begin{equation*}
  \hat P = \frac{1}{T}\sum_{t=1}^T(\hat w_t \otimes \hat w_{t-1}).
\end{equation*}
Then we have

\begin{thm}
  \label{thm:2}
  Let Assumptions \ref{assump_1} and \ref{assump_2} hold. If $N\geq cT^{1/r}$ for some constant $c$, then
  \begin{equation*}
    \mathbb{E}\norm{\hat Q - Q}^2 = O\left(T^{-1}\right),
  \end{equation*}
  and
  \begin{equation*}
    \mathbb{E}\norm{\hat P - P}^2 = O\left(T^{-1}\right)
  \end{equation*}
  as $T\to \infty$. Moreover, if $N>cT^{2/r}\ln^s T$ for some constants $c>0$ and $s>0$, then
  \begin{equation*}
    \norm{\hat Q - Q} = O\left(T^{-1/2}\ln^{1/2} T \right) \quad \text{a.s.}
  \end{equation*}
  and
  \begin{equation*}
    \norm{\hat P - P} = O\left(T^{-1/2}\ln^{1/2} T \right) \quad \text{a.s.}
  \end{equation*}
as $T\to \infty$.
\end{thm}

Theorem \ref{thm:2} shows that the sample analogue estimators $\hat Q$ and $\hat P$ based on the estimated density functions $(\hat f_t)$ are consistent, and have the same rates of convergence as those based on the true density functions $(f_t)$, which have been established in Theorem 4.1 and Corollary 4.1 of \cite{bosq-00}.

\subsection{Eigenvalues and Eigenvectors}
\label{sec:eigen_est}
The implementation of our methodology requires the estimation of eigenvalues and eigenvectors of $Q$, and $Q$ can be consistently estimated by $\hat Q$ by Theorem \ref{thm:2}. Naturally, the eigenvalues and eigenvectors of $Q$ are estimated by those of $\hat Q$, which we denote by $(\hat \lambda_k, \hat v_k)$. We assume that the estimated eigenvalues are distinct and order them so that $\hat \lambda_1> \hat \lambda_2 > \cdots$\footnote{We could potentially allow for multiplicity. However, in that case the eigenvectors could not be uniquely identified even after normalization. As a consequence we need to show consistency of eigenspaces, which complicates the notations and proofs substantially. In this paper, we assume for convenience that the eigenvalues are distinct.}. The eigenvalues and eigenvectors of $\hat Q$ are the pairs $(\hat \lambda, \hat v)$ such that $\hat Q\hat v = \hat \lambda \hat v$.

We may solve the eigen-problem either by discretizing $(\hat w_t)$ or by expanding $(\hat w_t)$ into linear combinations of an orthonormal basis of $H$ and obtaining the matrix representation of $\hat Q$ with respect to that basis. Either way, we transform the original problem into the problem of computing eigenvalues and eigenvectors of a matrix. Readers are referred to Section 8.4 in \cite{ramsey-silverman-05} for more details.

For definiteness, let us assume that the eigenspace corresponding to $(\lambda_k)$ is one dimensional for each $k$ from now on. This implies that Assumption \ref{assump_2}\ref{ass2_a} should be replaced by $\lambda_1>\lambda_2>\ldots>0$. We define $v_k'=\sgn\langle \hat v_k, v_k\rangle v_k$. Note that since both $(v_k)$ and $(-v_k)$ are eigenvectors corresponding to $\lambda_k$, the introduction of $(v_k')$ is essential for definitiveness of eigenvectors. The following lemma from \cite{bosq-00} associates $\abs{\hat \lambda_k - \lambda_k}$ and $\norm{\hat v_k - v_k'}$ with $\norm{\hat Q - Q}$.

\begin{lemma}
  \label{lemma:3}
  Let Assumptions \ref{assump_1} and \ref{assump_2}\ref{ass2_a} hold. We have
  \begin{equation*}
    \sup_{k\geq 1}\abs{\hat \lambda_k - \lambda_k} \leq \norm{\hat Q - Q}
  \end{equation*}
and
  \begin{equation*}
    \norm{\hat v_k - v_k'}\leq \tau_k \norm{\hat Q - Q},
  \end{equation*}
  where $\tau_1 = 2\sqrt{2}(\lambda_1 - \lambda_2)^{-1}$, and $\tau_k = 2\sqrt{2}\max\{(\lambda_{k-1} - \lambda_k)^{-1}, (\lambda_k - \lambda_{k+1})^{-1}\}$ for $k \geq 2$.
\end{lemma}

Let $\tau(K) = \sup_{1\leq k \leq K}(\lambda_k - \lambda_{k+1})^{-1}$ and set $K$ to be a function of $T$, i.e., $K_T$ such that $K_T\to \infty$ as $T\to \infty$. It follows directly from Theorem \ref{thm:2} and Lemma \ref{lemma:3} that

\begin{thm}
  \label{thm:4}
  Let Assumptions 1 and 2 hold. We have
  \begin{equation*}
    \sup_{k\geq 1} \abs{\hat \lambda_k - \lambda_k} = O\left( T^{-1/2} \ln^{1/2} T\right) \quad \text{a.s.}
  \end{equation*}
as $T\to\infty$.
\end{thm}

Our results in Theorem \ref{thm:4} shows that the convergence rates of the estimated eigenvalues, in the presence of estimation errors for the densities, are the same as in Theorem 4.4 and Corollary 4.3 in \cite{bosq-00}.

\subsection{Autoregression and Error Variance Operators}
Using the estimators introduced in Sections \ref{sec:variance_est} and \ref{sec:eigen_est}, we may consistently estimate the autoregressive operator $A$ by an estimator of $A_K$ given earlier in \eqref{eq:A_K}. To be explicit, we define
\begin{equation*}
  \hat Q_K^+ = \sum_{k=1}^K\hat \lambda_k^{-1}(\hat v_k \otimes \hat v_k)
\end{equation*}
and subsequently
\begin{equation}
  \label{eq:hat_A_K}
  \hat A_K = \hat P\hat Q_K^+.
\end{equation}

\begin{thm}
  \label{thm:5}
  Let Assumptions \ref{assump_1} and \ref{assump_2} hold. Moreover, suppose $N\geq cT^{2/r}\ln^s T$ for some constants $c>0$ and $s>0$, A is Hilbert-Schmidt, and $\frac{\ln T \left(\sum_{k=1}^K \tau_k\right)^2}{T\lambda_K^2} \to 0$ as $T\to \infty$. Then
  \begin{equation*}
    \norm{\hat A_K - A} \to 0 \quad \text{a.s.}
  \end{equation*}
as $T\to \infty$.
\end{thm}

Recall that $K$ is a function of the sample size $T$ such that $K\to \infty$ as $T\to \infty$. Theorem \ref{thm:5} establishes the consistency of $\hat A_K$. Consistency as in Theorem 8.8 in \cite{bosq-00} continues to hold in the presence of estimation errors for the densities.

Now we define
\begin{equation}
  \label{eq:1}
  \hat \Sigma = \frac{1}{T}\sum_{k=1}^T(\hat \varepsilon_t \otimes \hat \varepsilon_t),
\end{equation}
where $(\hat \varepsilon_t)$ obtained by
\begin{equation}
  \label{eq:residuals}
  \hat \varepsilon_t = \hat w_t - \hat A_K \hat w_{t-1}
\end{equation}
are the fitted residuals from the FAR in \eqref{eq:model} with $(w_t)$ replaced by $(\hat w_t)$. As is expected from Theorem \ref{thm:5}, we have

\begin{coro}
  \label{coro:6}
  Let the conditions in Theorem 5 hold. Then we have
  \begin{equation*}
    \norm{\hat \Sigma - \Sigma} \to 0 \quad \text{a.s.}
  \end{equation*}
as $T\to \infty$.
\end{coro}

Corollary \ref{coro:6} establishes the strong consistency of $\hat \Sigma$.

\section{Analysis of Distributional Autoregression}
This section provides some tools and asymptotics to analyze the distributional autoregression. In particular, we discuss two topics including the forecast of state density and the moment dynamics of state distributions.

\subsection{Forecast of State Density}
Our model can be used to obtain the forecasts of future state densities. For one-step forecast, we use
\begin{equation}
  \label{eq:one-step-forecast}
 \hat w_{T+1} = \hat A_K\hat w_{T},
\end{equation}
where $\hat A_K$ is the estimator of the autoregressive operator $A$ introduced in \eqref{eq:hat_A_K}. Multiple step forecasts can be obtained by recursively applying $\hat A_K$ to the forecasted values in the previous steps. Below we develop the asymptotic theory for our one-step forecast.

The development of our asymptotics require some extra technical conditions in addition to what we have in Assumptions \ref{assump_1} and \ref{assump_2}.
\begin{assump}
  \label{assump_3}
  We assume that
  \begin{enumerate}[label=(\alph*), nosep]
  \item \label{ass3_a} $\lambda_k$ is convex in $k$ for large $k$,
  \item \label{ass3_b} $\mathbb{E}\langle v_k, w_t\rangle^4 \leq M\lambda_k^2, k\geq 1$, for some constant $M>0$ independent of $k$, and,
  \end{enumerate}
\end{assump}

\noindent
Condition \ref{ass3_a} is weak and satisfied by many sequences including $\lambda_k = c/k^a$, $\lambda_k = c/(k^a\ln^b k)$ and $\lambda_k = ce^{-ak}$, where $a, b$ and $c$ are some positive constants. Let $\delta_k = \lambda_k - \lambda_{k+1}$. Then $\delta_k$ is a decreasing sequence by convexity of $\lambda_k$. The condition \ref{ass3_b} is met whenever the tail probability of $\langle v_k, w_t\rangle$ decreases fast enough. For instance, if $w_t$ is Gaussian, the condition holds with $M=3$. Note that $\langle v_k, w_t\rangle$ has mean zero and variance $\lambda_k$ for $k = 1, 2, \ldots$. It may not hold if the distribution of $\langle v_k, w_t\rangle$ for some $k$ has a thicker tail.

In the development of our subsequent theory, we denote  by $\Pi_K$ the projection onto the subspace spanned by the first $K$ principal eigenvectors $v_1, \ldots, v_K$ of $Q$, and denote by $\hat \Pi_k$ the projection onto the subspace spanned by the first $K$ principal eigenvectors $\hat v_1, \ldots, \hat v_K$ of $\hat Q$. The asymptotic theory that we develop for the one-step forecast is based on the foundational results due to \cite{mas-07}. To avoid technical difficulties, we follow \cite{mas-07} to compute the estimator $\hat A_K$ using samples only up to time $T-1$.

\begin{lemma}
  \label{lemma:7}
  Let Assumptions \ref{assump_1}, \ref{assump_2} and \ref{assump_3} hold. If
\begin{enumerate*}[label=(\alph*)]
\item $N\geq cT^{2/r}\ln^s T$ for some constants $c>0$ and $s>0$, \item \label{lemma7_ass_b} $\frac{1}{\sqrt{T}}K^{5/2}\ln^2 K\to 0$, and
\item \label{lemma7_ass_c} $\norm{Q^{-1/2}A}<\infty$,
\end{enumerate*}
then
  \begin{equation*}
    \sqrt{T/K}\left(\hat A_K\hat w_T - A\hat \Pi_K w_T\right) \to_d \mathbb{N}(0, \Sigma)
  \end{equation*}
as $T\to \infty$, where $\mathbb{N}(0, \Sigma)$ is a Gaussian random element taking values in the Hilbert space $H$, with mean zero and variance operator $\Sigma$.
\end{lemma}

\noindent
The condition \ref{lemma7_ass_b} requires that $K$ should not increase too fast relative to $T$. Loosely put, condition \ref{lemma7_ass_c} requires that $A$ should be at least as smooth as $Q^{1/2}$. If indeed they have common eigenvectors, then this condition holds if and only if the ordered sequence of eigenvalues of $A$ decays at least as fast as the ordered sequence of eigenvalues of $Q^{1/2}$.

The result in Lemma \ref{lemma:7} is not directly applicable to obtain the asymptotic confidence interval for the forecast of $(w_t)$. In particular, it has a random bias term $(A\hat \Pi_K -A) w_T$. This can be remedied, at the expense of a set of additional conditions as in the following theorem.

\begin{thm}
\label{thm:8}
Let the assumptions in Lemma \ref{lemma:7} hold. Suppose $\sup_{k\geq 1}\mathbb{E}\langle v_k, \Delta_t\rangle^4/\lambda_k^2 = O(N^{-2r})$, $(T/K)\sum_{k=K+1}^\infty \lambda_k\to 0$, $\sum_{i=1}^K\sum_{j=K+1}^\infty \lambda_i\lambda_j/(\lambda_i-\lambda_j)^2 = o(K)$, and that there exists constants $M$ and $\abs{b}<1$ such that $\sup_{v_p, v_m} \text{Corr}(\langle w_t, v_p\rangle \langle \Delta_t, v_m\rangle, \langle w_s, v_p\rangle \langle \Delta_s, v_m\rangle) \leq Mb^{\abs{s-t}}$. Then
\begin{equation*}
 \sqrt{T/K}\left(\hat A_K\hat w_T - Aw_T\right) \to_d \mathbb{N}(0, \Sigma)
\end{equation*}
as $T\to \infty$.
\end{thm}

\noindent
The newly introduced conditions in Theorem \ref{thm:8} can be restrictive, and depend upon the decaying rate of the eigenvalues of the variance operator $Q$, which is unobserved, and asymptotic independence of the process $(w_t\otimes \Delta_t)$. However, they are not prohibitively stringent. For instance, the conditions on the eigenvalues are satisfied for $\lambda_k = ce^{-k}$ where $c$ is some positive constant, and it seems that in many applications the eigenvalues indeed exhibit an exponential decay. The conditions on the correlations are satisfied if $(w_t\otimes \Delta_t)$ is a mixingale or $\rho$-mixing, with geometrically decaying mixing rates.  

The asymptotic result in Theorem \ref{thm:8} can be used to construct the asymptotic confidence interval for the one-step forecast of $(w_t)$. Indeed, we may easily deduce from the result in Theorem \ref{thm:8} that
\begin{equation*}
  \hat w_{T+1} - w_{T+1} = \hat A_K \hat w_T - Aw_T - \varepsilon_{T+1} \approx_d \mathbb{N}\left(0, \left( 1 + \frac{K}{T}\right)\Sigma\right).
\end{equation*}
It follows that for any $v\in H$ the interval forecast for $\langle v, w_{T+1}\rangle$ with confidence level $\alpha$ is given by
\begin{equation*}
  \left[\langle v, \hat w_{T+1}\rangle - z_{\alpha/2}\sqrt{(1+K/T)\langle v, \Sigma v\rangle}, \langle v, \hat w_{T+1}\rangle + z_{\alpha/2}\sqrt{(1+K/T)\langle v, \Sigma v\rangle}\right]
\end{equation*}
with $z_{\alpha/2}=\Phi^{-1}(1-\alpha/2)$, where $\Phi$ is the cumulative distribution function of the standard normal. Here $\Sigma$ could be replaced by its consistent estimate in implementation.

\subsection{Moment Dynamics of State Distributions}

In this section, we demonstrate how we may analyze the moment dynamics of state distributions. The analysis proves useful in learning the dependence structure in the moments of state distributions across time. For the analysis, we define a normalized moment basis $(u_k)$ in $H$ such that $u_k$ is the $k$-th order polynomial for $k=1,2,\ldots$, and that
\begin{enumerate}[label=(\alph*)]
\item \label{cond_1} $\langle 1,u_k\rangle = 0$ for all $k$, and
\item \label{cond_2} $\langle u_p, Qu_q\rangle = 1$ if $p=q$, and $0$ if $p\not =q$, for all $p$ and $q$.
\end{enumerate}
Such a basis could be obtained by the Gram-Schmidt orthogonalization procedure. For $C = [a,b]$, we may easily see that $u_1$ is given by $u_1(x) = C_1\big[x-(a+b)/2\big]$, where $C_1$ is a constant determined by $\langle u_1, Qu_1\rangle = 1$. Moreover, once $u_1, u_2, \ldots, u_p$ are obtained, we may readily find $u_{p+1}$ such that $\langle 1,u_{p+1}\rangle = 0$, $\langle u_q,Qu_{p+1} \rangle = 0$ for $q=1,\ldots,p$ and $\langle u_{p+1}, Q u_{p+1}\rangle = 1$. Note that $(u_k)$ is orthonormal with respect to the inner product $\langle \cdot, Q\cdot\rangle$ and generates $H$. That is, it is an orthonormal basis of $H$. Since $u_k$ is a polynomial of order $k$, we shall view it as the function that generates the normalized $k$-th order moment of state distributions.

Now we consider \eqref{eq:model_coor}, which defines the dynamics for the $v$-moment of the underlying state distributions. For any $v\in H$ given, write
\[
 A^\ast v = \sum_{k=1}^\infty \langle A^\ast v,Qu_k\rangle u_k
 = \sum_{k=1}^\infty \langle v,AQu_k\rangle u_k,
\]
so that we may deduce
\begin{equation}\label{mom-dec}
 \langle A^\ast v, w_{t-1}\rangle
 = \sum_{k=1}^\infty\langle v,AQu_k\rangle\langle u_k,w_{t-1}\rangle,
\quad
 \mathbb{E}\langle A^\ast v,w_{t-1}\rangle^2
 = \sum_{k=1}^\infty\langle v,AQu_k\rangle^2,
\end{equation}
which are quite useful to analyze the moment dynamics of state distributions. Note that we have
\[
 \mathbb E\langle u_p,w_t\rangle\langle u_q,w_t\rangle
 = \langle u_p,Qu_q\rangle
 = \left\{\begin{array}{cl} 1, & \mbox{if $p=q$}, \\ 0, & \mbox{if $p\ne q$}\end{array}\right.
\]
by construction. Consequently, for any $v\in H$ given, we have
\begin{equation}\label{mom}
 \mathbb{E}\langle v,w_t\rangle^2 = \mathbb{E}\langle A^\ast v,w_{t-1}\rangle^2
 + \mathbb{E}\langle v,\varepsilon_t\rangle^2
 = \sum_{k=1}^\infty\langle v,AQu_k\rangle^2 + \mathbb{E}\langle v,\varepsilon_t\rangle^2
\end{equation}
due to \eqref{eq:model_coor} and \eqref{mom-dec}.

In our analysis of the moment dynamics of state distributions, for any given $v\in H$ we define
\begin{equation}\label{r2}
 R_v^2 = 1-\frac{\mathbb{E}\varepsilon_t^2(v)}{\mathbb{E}\langle v,w_t\rangle^2}
 = 1-\frac{\langle v,\Sigma v\rangle}{\langle v,Qv\rangle}
\end{equation}
to be the $R^2$ of the $v$-moment of state distribution, and interpret the ratio
\begin{equation}\label{pik}
 \pi_v(k) = \frac{\mathbb{E}\langle v,AQu_k\rangle^2
 \langle u_k,w_{t-1}\rangle^2}{\mathbb{E}\langle v,w_t\rangle^2}
 = \frac{\langle v,AQu_k\rangle^2}{\langle v,Qv\rangle}
\end{equation}
as the proportion of the variance of the $v$-moment of the current state distribution contributed by the $k$-th moment of the past state distribution. Clearly, $R_v^2$ and $\pi_v(k)$ defined respectively in \eqref{r2} and \eqref{pik} can be consistently estimated by
\[
 \hat R_v^2 = 1-\frac{\langle v,\hat\Sigma v\rangle}{\langle v,\hat Qv\rangle}
\quad\mbox{and}\quad
 \hat\pi_v(k) = \frac{\langle v,\hat A_K\hat Qu_k\rangle^2}{\langle v,\hat Qv\rangle}
\]
for any $v\in H$, where $\hat\Sigma, \hat Q$ and $\hat A_K$ are the estimates of $\Sigma$, $Q$ and $A$, respectively, which have been introduced in Section 3.

\begin{table}[t]
\centering
\begin{tabular}{l@{\hskip 2em}l}
\hline \hline
Deviation  & Definition  \\ \hline
$D_{2}$ & $(\int (\hat f(x) - f(x))^2 dx)^{1/2}$ \\
$D_{1}$ & $\int |\hat f(x) - f(x)| dx$ \\
$D_{ks}$ & $ \sup_x |\hat F(x) - F(x)|$ \\
$D_{cm}$ & $\int (\hat F(x) - F(x))^2 dF(x)$ \\
$D_{m}$ & $|\int x\hat f(x)dx - \int xf(x)dx|$ \\
$D_{v}$ & $|\int [x - \int\hat f(x)dx]^2\hat f(x) dx - \int[x - \int f(x)dx]^2 f(x)dx|$ \\
\hline \hline
\end{tabular}
\caption{Forecast Error Measures}\label{tab:dev_measure}
\end{table}

\section{Empirical Applications}
In this section we use our methodology to analyze two financial markets. In the first, we study the intra-month distributions of the GBP/USD exchange rate 15-minute log returns. In the second, we study the cross-sectional distributions of monthly returns of stocks listed on the NYSE. We explore various aspects of the distributional dynamics in the foreign exchange and stock markets that would not be revealed if we only look at their aggregate time series.

We also present the forecast performances of our model. For each forecast period, we obtain the predicted density $\hat f$ and calculate how much it deviates from the actually observed density $f$, using the six measures given in Table \ref{tab:dev_measure}. $\hat F$ and $F$ in the table are the distribution functions corresponding to $\hat f$ and $f$, respectively. The first two measures, namely the $L^2$- and the $L^1$-deviations, are based on the density functions. The middle two measures, namely the Kolmogorov-Smirnov and the Cram\'{e}r-von Mises deviations, are based on the distribution functions. The last two measures, namely the (absolute) deviation in mean and in variance, are based on the first two moments of the densities. Notice that all these quantities are always nonnegative, and are zero when $\hat f = f$.

\subsection{Intra-Month Distributions of GBP/USD Exchange Rate 15-Minute Log Returns}

For a high frequency intra-month financial return series, we may consider the individual intra-month returns as draws from a common distribution, which varies by month, under the piecewise stationarity assumption. These draws may be dependent, hence accommodating the case in which the intra-month returns are not independent but are strictly stationary. This view not only provides a justification for applying our model to time-varying return distributions, but also reflects the fact that the intra-month returns exhibit statistical characteristics that do not carry over to longer horizons.

\begin{figure}
  \centering
  \includegraphics[width = \textwidth]{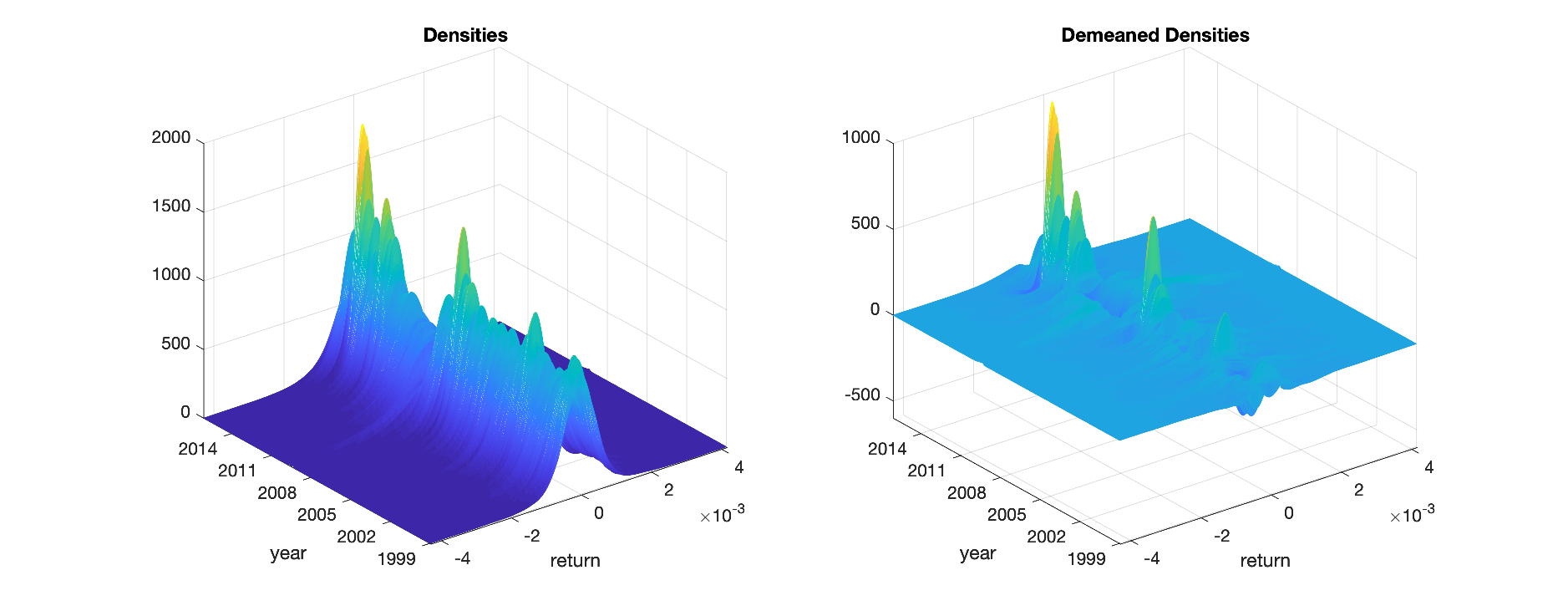}
  \caption{Intra-Month GBP/USD Exchange Rate Return Distributions}
\floatfoot{Notes: The left and right panels respectively present the densities and demeaned densities of the intra-month distributions obtained from the GBP/USD exchange rate 15-minute log returns.}
  \label{fig:dst_forex}
\end{figure}

In this application, we look at the intra-month distributions of the GBP/USD exchange rate 15-minute log returns. Since the foreign exchange market operates globally around-the-clock on weekdays, it is more natural to treat every four weeks as a month than to follow the calendar months. We thus set a month in this application to be a four week period. We use data from January 4th, 1999 to April 3rd, 2015, and  split the data into a total of 212 months using the four-week convention. The number of observations in each period ranges from 1550 to 1904, with a mean of 1880. We obtain the kernel density estimate for each period using the Epanechnikov kernel with the optimal bandwidth $h_t=2.3449\hat \sigma_t n_t^{-1/5}$, where $\hat \sigma_t$ and $n_t$ are respectively the sample standard deviation and the sample size in period $t$.
The support of the densities is set to be $[-0.0043, 0.0043]$, which includes more than $99.9\%$ of the actual observations. We plot the time series of both the densities and the demeaned densities in Figure \ref{fig:dst_forex}, from which we could easily see the time-varying nature of the corresponding distributions.

\begin{figure}[t]
  \centering
  \includegraphics[width = 0.8\textwidth]{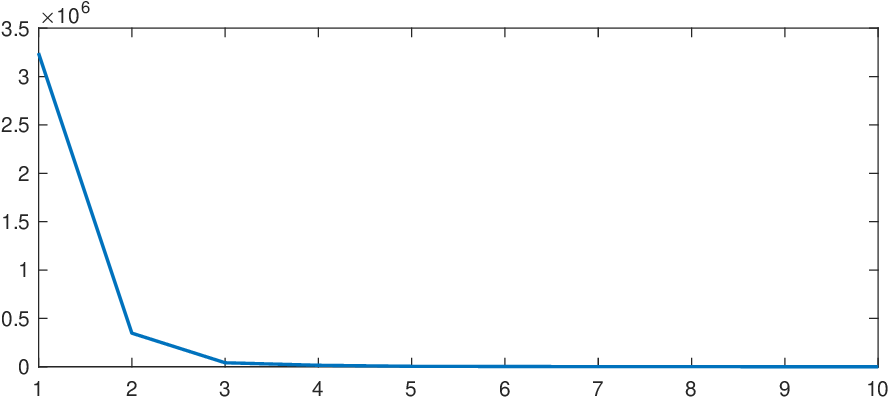}
  \caption{Scree Plot for Intra-Month GBP/USD Exchange Rate Return Distributions}
  \label{fig:scree_forex}
\floatfoot{Notes: The ten largest eigenvalues of the sample variance operator for the intra-month densities of the GBP/USD exchange rate 15-minute log returns are presented in descending order.}
\end{figure}

To implement our method, we represent the densities with the Daubechies wavelets using 1037 basis functions. Each density is then represented as a 1037-dimensional vector whose coordinates are the wavelet coefficients of the density function. Thus we transform the functional time series into a time series of high dimensional vectors. We obtain the matrix representations of the sample variance operator $\hat Q$ and the sample first-order autocovariance operator $\hat P$ with respect to that basis, which are respectively the sample covariance and the sample first-order autocovariance matrices of the vector time series. We then estimate the eigenvalues and eigenvectors of the variance operator $Q$ respectively by the eigenvalues and eigenvectors of the sample covariance matrix of the vector time series. When we approximate the inverse of the variance operator, we set $K=4$ to obtain the best rolling out-of-sample forecast performance (see below). This is the data-driven method we propose for choosing the value of the nuisance parameter $K$. As another justification for our choice of $K$ from the perspective of functional principal component analysis, we note that the first 4 principal components together explain 99.7\% of the total variation in the density process. See Figure \ref{fig:scree_forex} for the scree plot of the ten largest eigenvalues of the variance operator.

\begin{figure}[!t]
  \centering
  \includegraphics[width = 0.8\textwidth]{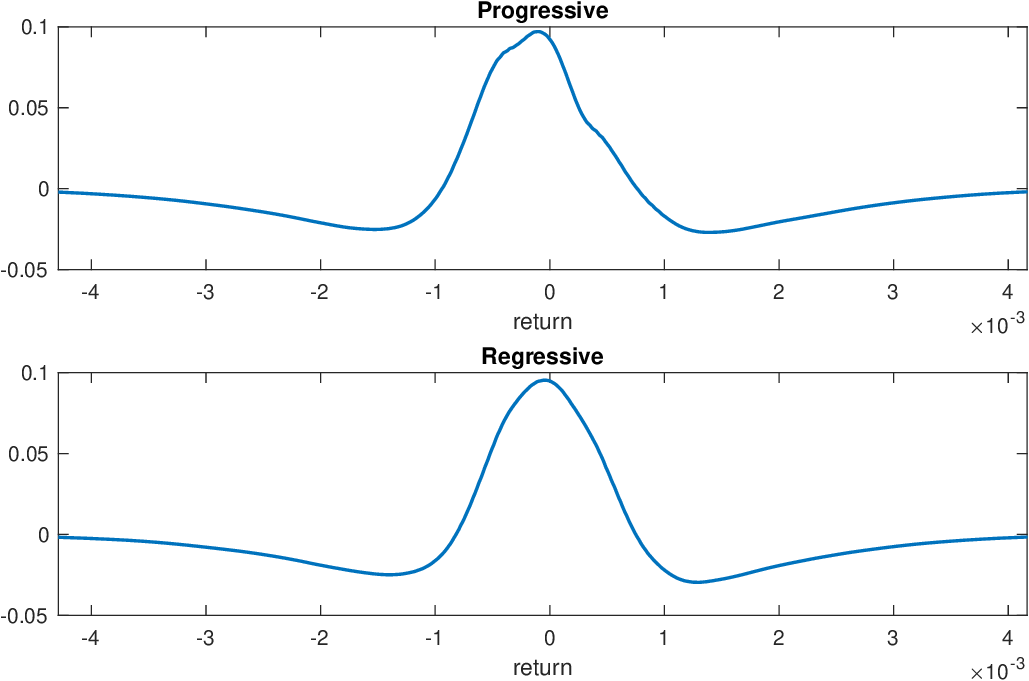}
  \caption{Leading Features of Intra-Month GBP/USD Exchange Rate Return Distributions}
\floatfoot{Notes: The upper and lower panels present respectively the leading progressive and regressive features of the intra-month distributions of the GBP/USD exchange rate 15-minute log returns.}\vspace{-0.1in}
  \label{fig:input_output_forex}
\end{figure}

As an illustration of how we may use the tools developed in this paper to perform interesting analyses, we first obtain the leading progressive and regressive features for this density process. Each feature is represented by its values at 1024 points evenly spaced on the support of the distributions. See Figure \ref{fig:input_output_forex}. The leading progressive feature is a concentrating feature: If it is loaded with a positive scale, the distribution will be more concentrated around its center. It also shows that the relative frequency of observations with small returns is most important for determining distributions in the future. The leading regressive feature shows it is also the relative frequency of observations with small returns that is most affected by distributions in the past. Both features indicate that the observations with extreme returns are not important in the distributional dynamics.

\begin{figure}[t]
  \centering
  \includegraphics[width = \textwidth]{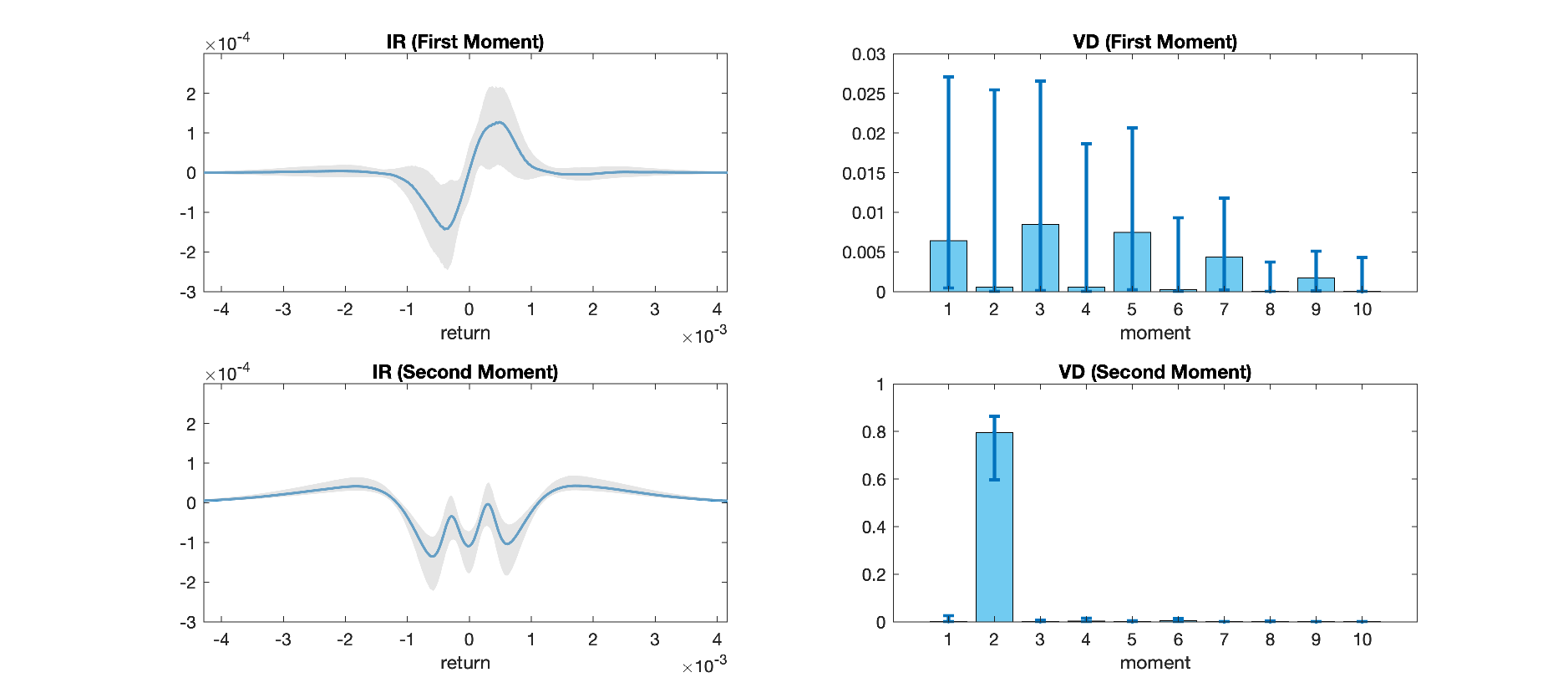}
  \caption{Impulse Responses and Variance Decompositions for Moments of Intra-Month GBP/USD Exchange Rate Return Distributions}
\floatfoot{Notes: The left two panels respectively plot the impulse responses of the first moment (upper panel) and the second moment (lower panel) of the current distribution of the GBP/USD exchange rate 15-minute log returns to Dirac-$\delta$ impulses to the last month's density function at different levels of return. The shaded areas give the corresponding 95\% residual bootstrap confidence bands based on 2000 repetitions. The right two panels depict respectively the proportions of variances of the first moment (upper panel) and the second moment (lower panel) of the  current distribution of the GBP/USD exchange rate 15-minute log returns that are explained by the variances of the first ten moments of the last month's distribution, with their 95\% residual bootstrap confidence intervals.}
  \label{fig:var_decomp_forex}\vspace{-0.1in}
\end{figure}

We also illustrate how various aspects in the current distribution respond to shocks to the past distribution. The left two panels in Figure \ref{fig:var_decomp_forex} present the current first two moments' responses with respect to Dirac-$\delta$ impulses to the previous distribution at different levels of return. The shaded areas give the corresponding 95\% residual bootstrap confidence bands based on 2000 repetitions. The top left panel shows that the mean of the current distribution is most affected by shocks to the relative frequencies of the observations with small-to-moderate returns in the previous period, and shocks to the relative frequencies of the observations with large or extreme returns in the previous period are not important. Moreover, we see the momentum effect, i.e., a larger number of observations with small-to-moderate negative (positive) returns in the previous period is likely to result in a lower (higher) first moment in the current period. The bottom left panel shows that the second moment of the current distribution, on the other hand, is affected by shocks to the relative frequencies of returns at all levels in the previous period. One may expect a smaller second moment in the current period if there were more observations with small-to-moderate returns in the previous period, and a larger second moment in the current period if there were more observations with large returns in the previous period. The effects of small-to-moderate returns in the previous period on the second moment in the current period are not monotonic. The stabilizing effects of some moderate returns are at least as strong as those of the small returns.

\begin{figure}[t]
  \centering
  \includegraphics[width = \textwidth]{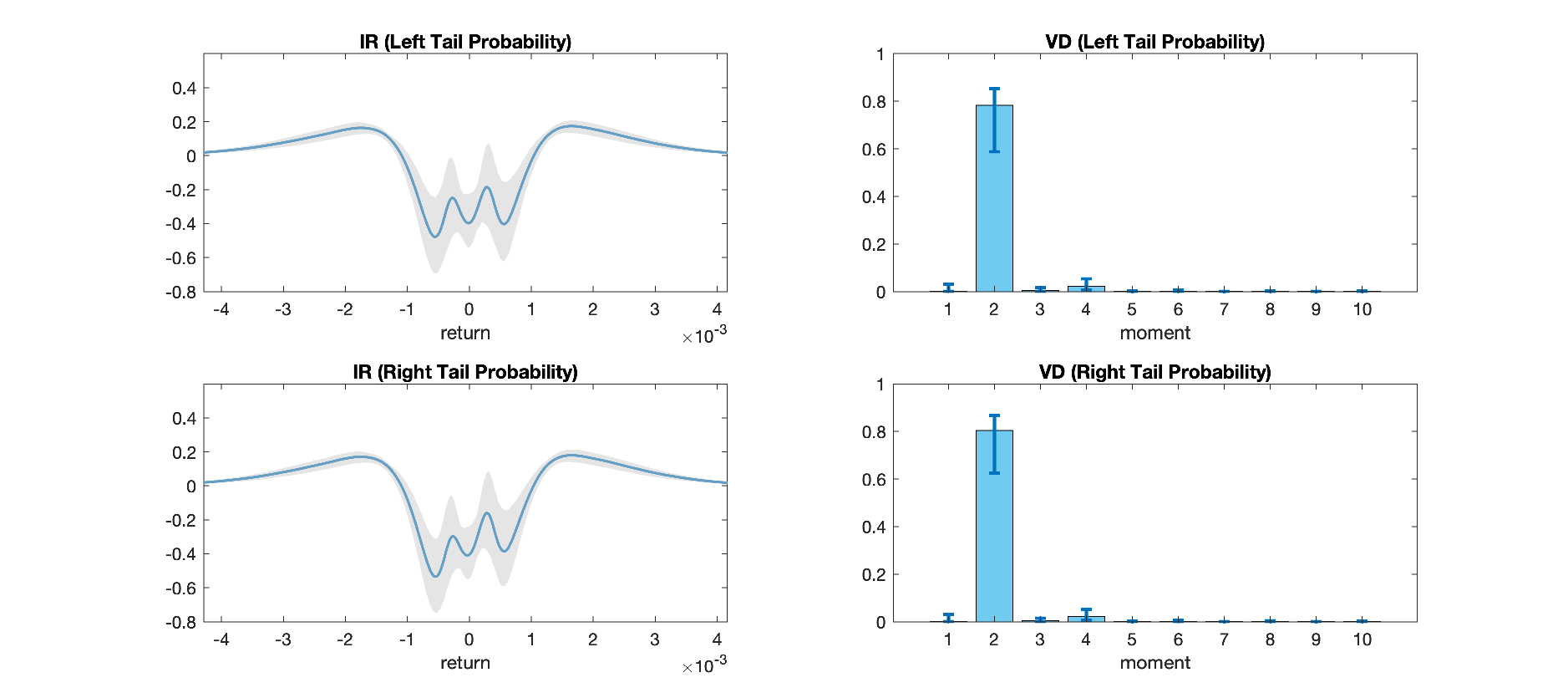}
  \caption{Impulse Responses and Variance Decompositions for Tail Probabilities of Intra-Month GBP/USD Exchange Rate Return Distributions}
  \label{fig:tail_prob_forex}
\floatfoot{Notes: The left two panels respectively plot the impulse responses of the left tail probability (upper panel) and the right tail probability (lower panel) of the current distribution of the GBP/USD exchange rate 15-minute log returns to Dirac-$\delta$ impulses to the last month's density function at different levels of return. The shaded areas give the corresponding 95\% residual bootstrap confidence bands based on 2000 repetitions. The right two panels depict respectively the proportions of variances of the left tail probability (upper panel) and the right tail probability (lower panel) of the  current distribution of the GBP/USD exchange rate 15-minute log returns that are explained by the variances of the first ten moments of the last month's distribution, with their 95\% residual bootstrap confidence intervals.}\vspace{-0.1in}
\end{figure}

The right two panels give the proportions of the variances of the current first and second moments that are explained by the variances of the first ten moments of the previous distribution, with their 95\% residual bootstrap confidence intervals. The corresponding $R_v^2$ are respectively $0.0380$ and $0.8091$, with 95\% bootstrap confidence intervals $[0.0085, 0.1186]$ and $[0.6214, 0.8727]$. Though many integral moments, especially those of odd orders, affect the current mean, their overall effect is almost negligible. In contrast, the past second moment, whose variance explains about $80\%$ of the total variance, is very informative about the current second moment, while the other past moments provide almost no information. Our findings here are entirely consistent with the widespread and repeated observations by many empirical researchers that in financial markets the mean is usually difficult to predict, while the volatility process is much more persistent and therefore much more predictable.

Figure \ref{fig:tail_prob_forex} gives the impulse responses and the variance decompositions of the tail probabilities. The left tail includes the returns that are smaller than the 5th percentile of the mean density, and the right tail includes the returns that are greater than the 95th percentile of the mean density. In this application, the responses of the two tail probabilities to shocks in the past density are quite symmetric. Shocks to, and subsequent changes in, the relative frequencies of returns at all levels in the previous period matter. A larger number of observations with small-to-moderate returns in the last period tends to presage thinner tails of the current distribution, while a larger number of observations with large returns in the previous period tends to presage thicker tails of the current distribution. The tail responses of small-to-moderate returns are non-monotonic, and are essentially the same as their responses to the second moment. In the variance decomposition analysis, about $80\%$ of the variances of the current tail probabilities are explained by the variances of the past second moment, which shows that the past second moment is very informative about the current tail probabilities.

Finally, to evaluate the prediction performance of our model in this application, we make rolling out-of-sample forecasts of the state densities for the last 50 periods. For each forecast period, we use the data prior to that period to estimate the autoregressive coefficient. We dynamically choose $K$ by cross-validation, using the five periods before the forecast period as the validation periods. We then make one-period-ahead forecasts of the demeaned density as in \eqref{eq:one-step-forecast}. We add the mean density to the forecasted demeaned density to obtain the density forecast. We normalize the forecasted density by setting its negative values to zero and then rescaling it so that it integrates to one. In the end, we calculate the forecast error. For comparison, we consider two other predictors,  which are respectively the time average of all the estimated densities and the estimated density of the last period. We label our functional autoregressive predictor as the FAR predictor, and the other two  respectively as the AVE predictor and the LAST predictor. The AVE and the LAST predictors are used as benchmarks: If the true process of the random densities is indeed independent and identically distributed over time, one would expect the AVE predictor to perform well. If the true process is a nonstationary martingale process, one would expect the LAST predictor to perform the best. Comparisons with the two benchmarks reveal information on how far away the actual process is from the i.i.d. scenario and from the nonstationarity scenario.

\begin{table}[t]
\centering
\begin{tabular}{l@{\hskip 1.5em}l@{\hskip 1.5em}l@{\hskip 1.5em}l}
\hline \hline
& FAR & AVE & LAST \\ \hline
$D_{2}$ & $2.67(2.15)$ & $5.54(4.19)$ & $2.67(2.12)$ \\
$D_{1} (10^{-1})$ & $1.14(0.97)$ & $2.49(2.04)$ & $1.15(0.90)$ \\
$D_{ks} (10^{-2})$ & $3.10(2.94)$ & $6.47(5.53)$ & $3.28(2.73)$ \\
$D_{cm} (10^{-3})$ & $0.65(0.35)$ & $2.75(1.26)$ & $0.68(0.27)$ \\
$D_{m} (10^{-5})$ & $0.82(0.73)$ & $0.81(0.71)$ & $1.27(1.14)$ \\
$D_{v} (10^{-7})$ & $0.60(0.40)$ & $1.78(1.87)$ & $0.58(0.51)$ \\
\hline \hline
\end{tabular}
\caption{Forecast Errors for Intra-Month GBP/USD Exchange Rate Return Distributions}\label{tab:est_forex}
\end{table}

Table \ref{tab:est_forex} reports the means and medians (in parentheses) of the 50 periods of forecast errors for each of the three predictors. In terms of the first four measures, the FAR predictor performs roughly the same as the LAST predictor, and they outperform the AVE estimator significantly. This suggests that the true process does not resemble an i.i.d. process, and that there is likely to be strong persistency in the density process. In terms of predicting the mean, the FAR predictor performs about the same as the AVE predictor, and much better than the LAST predictor.  In terms of predicting the variance, the FAR predictor performs about the same as the LAST predictor, and much better than the AVE estimator. These results corroborate the findings that in financial markets the mean is usually difficult to predict, while the volatility is quite persistent, and therefore much more predictable.

\subsection{Cross-Sectional Distributions of NYSE Stocks Monthly Returns}

\begin{figure}
  \centering
  \includegraphics[width = \textwidth]{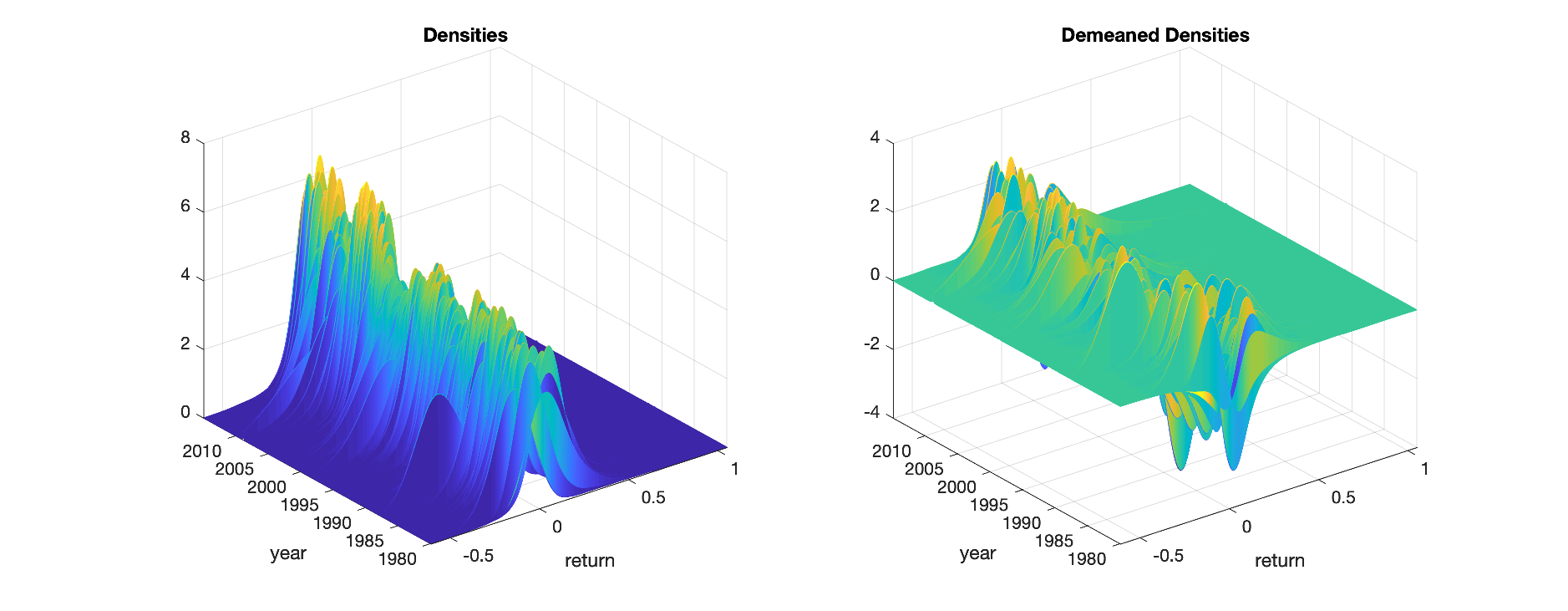}
  \caption{Cross-Sectional Distributions of NYSE Stocks Monthly Returns}
\floatfoot{Notes: The left and right panels respectively present the densities and demeaned densities of the  cross-sectional distributions obtained from monthly returns of stocks listed on the New York Stock Exchange.}
  \label{fig:dst_nyse}
\end{figure}

In this application, we look at the cross-sectional distributions of monthly returns of stocks listed on the New York Stock Exchange. In each month, the individual stock returns are viewed as draws from a common distribution. This distribution changes from month to month and is assumed to follow our FAR model \eqref{eq:model}. We use data from January of 1980 to December of 2014, a total of 420 months. In each month, we only take into account the stocks that are traded. The number of observations each month ranges from 1926 to 3076, with a mean of 2464. We use the same method as in the previous application to estimate the densities. The support of the density in this application is set to be $[-0.6071, 1.0548]$, which includes more than $99.9\%$ of the actual observations. Figure \ref{fig:dst_nyse} plots the times series of the densities and the demeaned densities of the NYSE stocks monthly returns.

\begin{figure}[t]
  \centering
  \includegraphics[width = 0.8\textwidth]{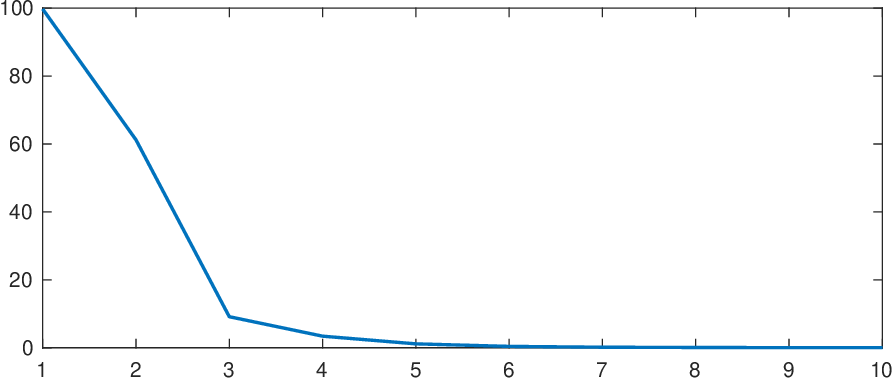}
  \caption{Scree Plot for Cross-Sectional Distributions of NYSE Stocks Monthly Returns}
\floatfoot{Notes: The ten largest eigenvalues of the sample variance operator for the cross-sectional densities of the NYSE stocks monthly returns are presented in descending order.}
  \label{fig:scree_nyse}
\end{figure}

\begin{figure}[!h]
  \centering
  \includegraphics[width = 0.8\textwidth]{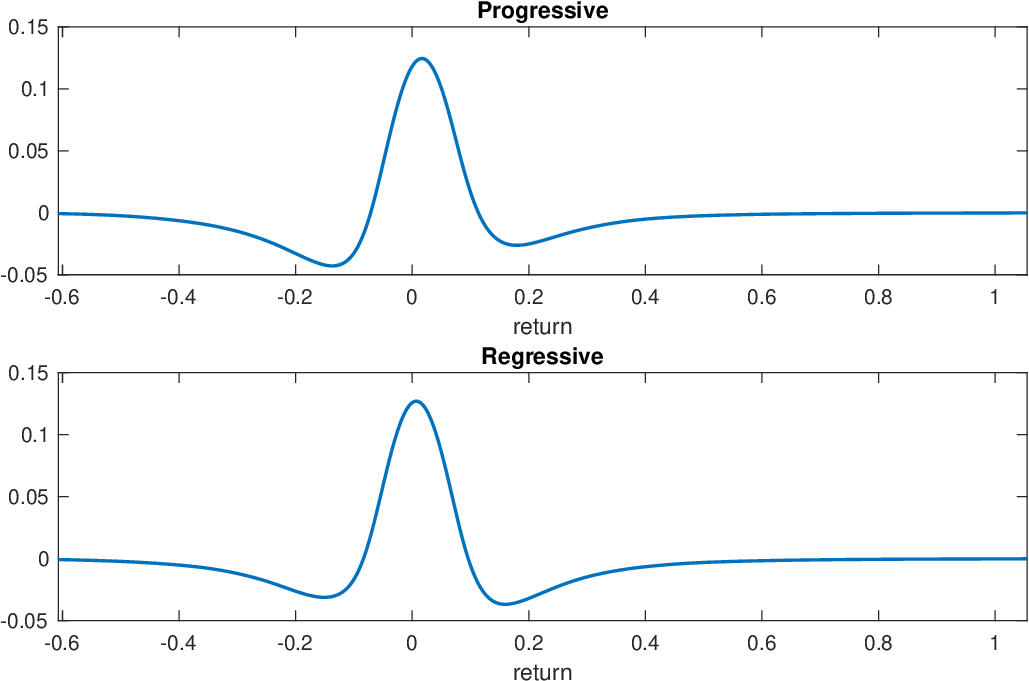}
  \caption{Leading Features of Cross-Sectional Distributions of NYSE Stocks Monthly Returns}
\floatfoot{Notes: The upper and lower panels respectively present the leading progressive and regressive features of the cross-sectional distributions of the NYSE stocks monthly returns.}
  \label{fig:input_output_nyse}\vspace{-0.1in}
\end{figure}

We use the same procedures as in the previous application for model estimation. We set $K=3$ based on the out-of-sample forecast performance. Figure \ref{fig:scree_nyse} gives the scree plot of the eigenvalues of the sample variance operator in this application. The first 3 principal components explain $97.0\%$ of the total variation in the density process. Figure \ref{fig:input_output_nyse} demonstrates the leading progressive and regressive features for the density process of the NYSE stocks monthly returns. As in the GBP/USD exchange rate application, it is the relative frequency of small returns that is most critical in the distribution dynamics. The small returns play a central role in affecting the future distribution, and they are most affected by the distribution in the previous period.

\begin{figure}[!h]
  \centering
  \includegraphics[width = \textwidth]{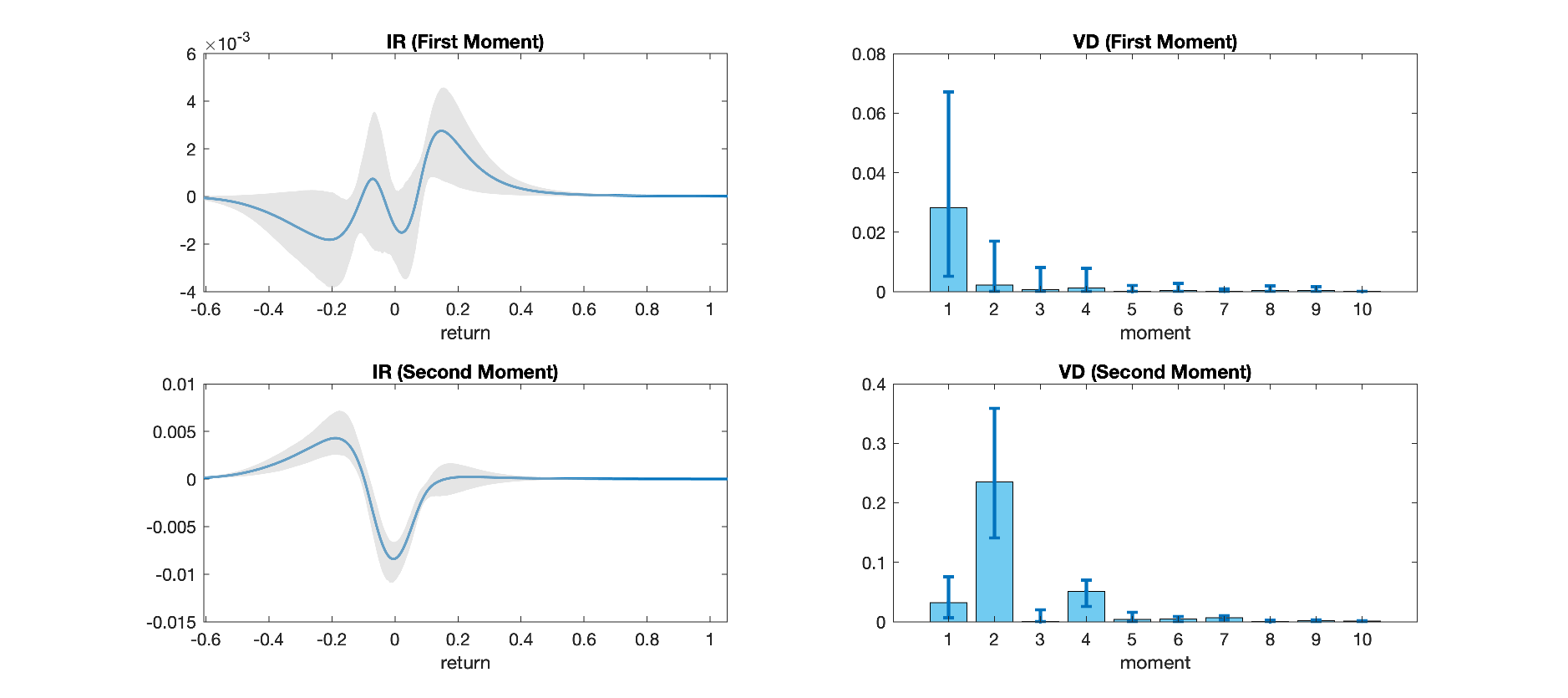}
  \caption{Impulse Responses and Variance Decompositions for Moments of Cross-Sectional Distributions of NYSE Stocks Monthly Returns}
  \label{fig:var_decomp_nyse}
\floatfoot{Notes: The left two panels respectively plot the impulse responses of the first moment (upper panel) and the second moment (lower panel) of the current distribution of the NYSE stocks monthly returns to Dirac-$\delta$ impulses to the last month's density function at different levels of return. The shaded areas give the corresponding 95\% residual bootstrap confidence bands based on 2000 repetitions. The right two panels depict respectively the proportions of variances in the first moment (upper panel) and in the second moment (lower panel) of the  current distribution of the NYSE stocks monthly returns that are explained by the variances of the first ten moments of the last month's distribution, with their 95\% residual bootstrap confidence intervals.}\vspace{-0.1in}
\end{figure}

In Figure \ref{fig:var_decomp_nyse}, we provide the impulse responses and the variance decompositions of the first two moments in this application. The corresponding $R_v^2$ are $0.0368$ and $0.3376$ respectively with 95\% residual bootstrap confidence intervals $[0.0108, 0.0815]$ and $[0.2199, 0.4709]$, which implies that the first moment is much less predictable than the second moment through the integral moments of the past distribution. Unlike the exchange rate returns application, in which the current mean responds only to changes in the relative frequencies of small returns in the previous period, the current mean in this application responds more significantly to changes in the relative frequencies of moderate-to-large returns in the previous period. The typical momentum effect is generated by stocks with moderately large returns, both positive and negative. For moderately small negative returns, we even find some evidence of a reverse momentum effect, a pattern which is also found in international stock returns by \citet{fama-french-12}. As expected, the second moment of current return distribution decreases as we have more small return stocks in the past distribution. Interestingly, the current second moment responds positively to changes in the relative frequencies of moderate-to-large negative returns, but not to changes in the relative frequencies of positive returns, in the previous period. This is presumably due to the leverage effect, and our result here suggests that the leverage effect is asymmetric, being effective only for negative return shocks. The past integral moments are not informative about the current mean. They are relatively more informative about the current second moment. However, the relevance is smaller than in the previous application, and the past fourth and first moments appear to play some non-negligible roles.

\begin{figure}[!h]
  \centering
  \includegraphics[width = \textwidth]{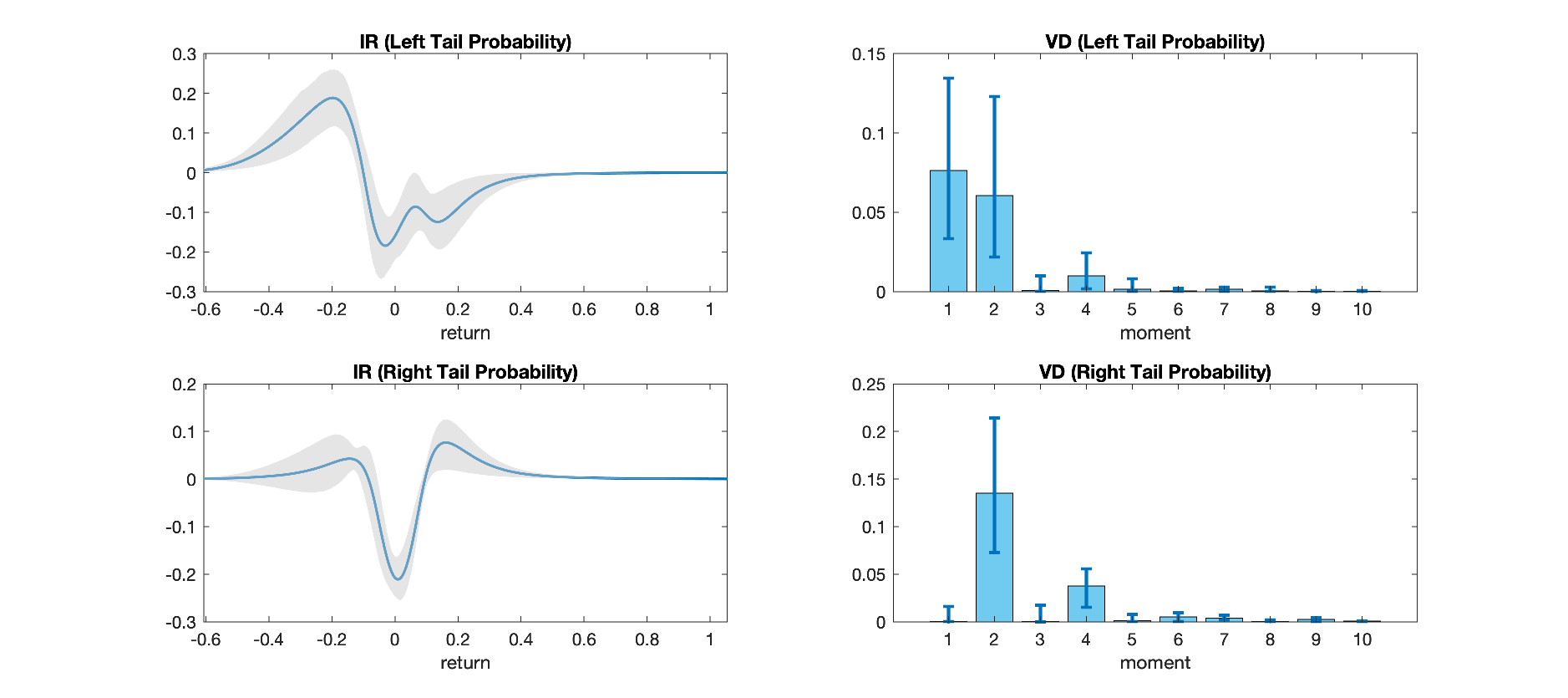}
  \caption{Impulse Responses and Variance Decompositions for Tail Probabilities of Cross-Sectional Distributions of NYSE Stocks Monthly Returns}
  \label{fig:tail_prob_nyse}
\floatfoot{Notes: The left two panels respectively plot the impulse responses of the left tail probability (upper panel) and the right tail probability (lower panel) of the current distribution of the NYSE stocks monthly returns to Dirac-$\delta$ impulses to the last month's density function at different levels of return. The shaded areas give the corresponding 95\% residual bootstrap confidence bands based on 2000 repetitions. The right two panels depict respectively the proportions of variances in the left tail probability (upper panel) and in the right tail probability (lower panel) of the  current distribution of the NYSE stocks monthly returns that are explained by the variances of the first ten moments of the last month's distribution, with their 95\% residual bootstrap confidence intervals.}\vspace{-0.1in}
\end{figure}

Figure \ref{fig:tail_prob_nyse} presents the impulse responses and the variance decompositions of the tail probabilities in this application. The tail events are defined in the same way as in the last application. In contrast with the previous application, the left tail probabilities and the right tail probabilities in this application respond differently to Dirac-$\delta$ impulses to the last period's density. As is expected, positive shocks to the relative frequencies of small returns decrease both the left and the right tail probabilities in the next period. Positive shocks to the relative frequencies of moderate-to-large negative returns tend to presage a thicker left tail, whereas positive shocks to the relative frequencies of moderate-to-large positive returns tend to presage a thicker right tail. Of the two effects, the former appears to be much more significant than the latter. Moreover, positive shocks to the relative frequencies of moderate-to-large positive returns decrease the left tail probability, whereas positive shocks to the relative frequencies of moderate-to-large negative returns also increase the right tail probability, though less substantially. On the other hand, the current left tail probability is related to both the past first and second moments, while the current right tail probability hinges more on the past second and fourth moments.

In this application, we make rolling out-of-sample forecasts for the last 100 periods since we have longer time series available. Table \ref{tab:est_nyse} summarizes the prediction results. In terms of the first four measures, our FAR predictor performs the best while the LAST estimator performs the worst. In terms of predicting the mean, the FAR predictor performs slightly worse than the AVE predictor, and much better than the LAST predictor. In terms of predicting the variance, the FAR predictor performs worse than the LAST predictor, but much better than the AVE predictor. These results suggest that this density process is likely to be stationary, and our model has good forecast power. However, there may be some particular directions in which the process is close to nonstationary, as volatility appears to be quite persistent.

\begin{table}[!t]
\centering
\begin{tabular}{l@{\hskip 1.5em}l@{\hskip 1.5em}l@{\hskip 1.5em}l}
\hline \hline
& FAR & AVE & LAST \\ \hline
$D_{2} (10^{-1})$ & $5.17(4.79)$ & $5.95(5.95)$ & $6.55(5.16)$ \\
$D_{1} (10^{-1})$ & $3.44(2.92)$ & $3.93(3.73)$ & $4.29(3.57)$ \\
$D_{ks} (10^{-1})$ & $1.62(1.42)$ & $1.75(1.46)$ & $2.10(1.79)$ \\
$D_{cm} (10^{-2})$ & $1.91(1.03)$ & $1.94(0.94)$ & $3.31(1.64)$ \\
$D_{m} (10^{-2})$ & $4.05(2.73)$ & $3.90(2.59)$ & $5.25(3.73)$ \\
$D_{v} (10^{-3})$ & $4.04(3.30)$ & $7.10(6.47)$ & $2.89(1.79)$ \\
\hline \hline
\end{tabular}
\caption{Forecast Errors for Cross-Sectional Distributions of NYSE Stocks Monthly Returns}\label{tab:est_nyse}
\end{table}

\section{Simulations}
In this section we study the finite sample properties of our FAR estimator by conducting out-of-sample forecast using simulated data. To make our simulations more realistic and practically relevant, we use the estimation results from the previous two empirical applications and generate data that mimics the estimated empirical models as closely as possible.

To be specific, for each application, we take the estimated coefficient $\hat A_K$ to be the autoregressive coefficient in the FAR(1) data generating process given in \eqref{eq:model} for the process $(w_t)$. In each iteration of simulations, the error process is bootstrapped from the demeaned residuals of the estimated FAR(1) model, given in \eqref{eq:residuals}. Once we obtain the simulated demeaned densities $(w_t)$, we add back the time average of the estimated densities $\bar f$ and normalize them as in the forecast procedures in the two applications to obtain simulated density functions. In each simulation, we start from $\varepsilon_0 = 0$, simulate 1000 periods and take the last $T+1$ periods. The actual values for $T$ will be specified below. We take the time series of the last $T+1$ density functions as the true process of densities. We then simulate $N$ observations from the density in each period using acceptance (or rejection) sampling. The actual values for $N$ will be specified below. Acceptance sampling is a widely used sampling method, especially in Bayesian statistics, when one can not directly sample from the distribution but the corresponding density function is known. The researcher samples from a proposed distribution which is different from the target distribution, and accepts or rejects each draw with a probability that is determined by the ratio of the proposal and target density functions evaluated at the draw. In our sampling procedure, we use the uniform distribution on the support as the proposal distribution. We take the accepted draws as the simulated observations.

We implement our functional autoregression using the simulated observations from the first $T$ periods, leaving the last period for forecast performance evaluation. The estimation procedure as well as the parameter settings are the same as in the empirical applications, except that in the kernel density estimation step, we use the Normal kernel instead of the Epanechnikov kernel, with the corresponding optimal bandwidth. The Normal kernel performs better in our simulations in that it gives smaller kernel density estimation error. Once we obtain the estimated autoregressive coefficient, we do one-step-ahead forecast for the last period and compare our forecast with the true density in that period. The forecast procedure is exactly the same as in the two applications in Section 5. We evaluate the difference of our forecast from the true density using the six criteria introduced in the previous section.

We set $T$ to be 50, 100, 200, or 500, and $N$ to be 100, 200, 500 or 5000. For each pair of $(T, N)$, the mean and median of the forecast errors based on 2000 iterations of simulations are reported in Table \ref{tab:simu_forex} and Table \ref{tab:simu_nyse}. For convenience, we label the simulations based on the two estimated models as the ``FOREX simulations'' and the ``NYSE simulations'' respectively. We make the following observations.

First, for both sets of simulations, the forecast performance is not very sensitive to the number of periods ($T$) we have in the estimation. Once we have at least 100 periods, results become highly stable. It seems that a total of 50 periods is still acceptable, although the forecast errors are slightly bigger.

Second, forecast performances are more sensitive to the number of observations ($N$) in each period. For both sets of simulations and for all values of $T$, when we increase the number of observations in each period, the forecast errors decrease non-negligibly. This suggests that a large part of the forecast error comes from the density estimation procedure. This agrees with the usual observation that kernel density estimation converges at a relatively slow rate.

Third, for both sets of simulations, and for all combinations of $T$ and $N$, the FAR estimator performs best in terms of the first four criteria. This implies that our method has good curve fitting property, in terms of both the density function (the first two criteria) and the distribution function (the middle two criteria).
In the FOREX simulations, the FAR predictor performs roughly the same as the AVE predictor, and much better than the LAST estimator in predicting the mean. In predicting the variance, the FAR predictor performs roughly the same as the LAST predictor, and much better than the AVE predictor. In the NYSE simulations, the FAR predictor has the best performance in predicting both the mean and the variance.

To summarize, we find that the FAR predictor forecasts densities accurately, and that the forecast is stable, even with a relatively short time series.

\section{Conclusions}
This paper presents a model for time-varying densities that follow an autoregressive stationary process in some function space. We explore the properties and implications of this model, showing how it relates to many familiar time series models such as the ARCH-type models. Our approach is highly nonparametric, in that it does not assume any parametric form for the underlying distributions. We have outlined the procedures to estimate the model assuming that the densities are not observable, and have shown the consistency of the estimator as both the sample size for the density estimation and the time horizon go to infinity. Furthermore, we illustrate how to forecast densities using our model, and establish the asymptotic normality of the predictor. We develop tools such as progressive/regressive analysis, impulse response analysis and variance decomposition to help study the dynamics in various aspects of the distribution process. We discover several interesting features in the distribution process of the intra-month GBP/USD exchange rate 15-minute log returns and in the distribution process of the NYSE stocks monthly returns. Finally, Monte Carlo simulations show that the FAR density predictor outperforms  its competitors. In sum, the FAR modeling for time series of state distributions is a very general framework that is highly flexible, readily applicable, and it provides a powerful method to extract information of a distributional process and generate accurate forecasts.

\begin{landscape}
\begin{table}
\tiny\centering
\begin{tabular}{ll|lll|lll|lll|lll}
\hline \hline
 & & $ N = 100 $ & & & $ N = 200 $ & & & $ N = 500 $ & & & $ N = 5000 $ & &  \\ \hline
 & & FAR & AVE & LAST& FAR & AVE & LAST& FAR & AVE & LAST& FAR & AVE & LAST\\ \hline
$ T = 50 $ & $D_{2}$ & $\bm{3.37}(3.00)$ & $4.46(3.87)$ & $4.10(3.85)$  & $\bm{3.03}(2.57)$ & $4.46(3.81)$ & $3.47(3.11)$  & $\bm{2.62}(2.21)$ & $4.27(3.62)$ & $2.89(2.52)$  & $\bm{2.12}(1.74)$ & $4.13(3.54)$ & $2.15(1.73)$ \\
 & $D_{1} (10^{-1})$ & $\bm{1.63}(1.47)$ & $2.15(1.89)$ & $2.02(1.92)$  & $\bm{1.44}(1.26)$ & $2.14(1.88)$ & $1.70(1.56)$  & $\bm{1.25}(1.08)$ & $2.06(1.80)$ & $1.40(1.24)$  & $\bm{1.01}(0.84)$ & $1.99(1.75)$ & $1.03(0.85)$ \\
 & $D_{ks} (10^{-2})$ & $\bm{4.48}(4.03)$ & $5.65(4.98)$ & $6.27(5.94)$  & $\bm{3.90}(3.43)$ & $5.56(4.91)$ & $5.08(4.62)$  & $\bm{3.37}(2.89)$ & $5.36(4.69)$ & $4.07(3.70)$  & $\bm{2.79}(2.41)$ & $5.19(4.49)$ & $2.92(2.50)$ \\
 & $D_{cm} (10^{-3})$ & $\bm{0.98}(0.57)$ & $1.88(0.98)$ & $1.79(1.30)$  & $\bm{0.76}(0.41)$ & $1.85(0.96)$ & $1.17(0.78)$  & $\bm{0.58}(0.30)$ & $1.72(0.88)$ & $0.76(0.48)$  & $\bm{0.42}(0.20)$ & $1.62(0.82)$ & $0.43(0.22)$ \\
 & $D_{m} (10^{-5})$ & $1.98(1.61)$ & $\bm{1.23}(1.03)$ & $5.31(4.12)$  & $1.61(1.31)$ & $\bm{1.10}(0.91)$ & $3.96(3.19)$  & $1.24(0.99)$ & $\bm{1.04}(0.87)$ & $2.71(2.15)$  & $1.06(0.86)$ & $\bm{1.01}(0.84)$ & $1.48(1.16)$ \\
 & $D_{v} (10^{-7})$ & $1.38(1.11)$ & $2.07(1.78)$ & $\bm{1.24}(0.85)$  & $1.17(0.89)$ & $2.03(1.77)$ & $\bm{1.04}(0.69)$  & $1.07(0.85)$ & $2.03(1.77)$ & $\bm{0.90}(0.62)$  & $0.82(0.58)$ & $1.99(1.71)$ & $\bm{0.76}(0.50)$ \\
\\
$ T = 100 $ & $D_{2}$ & $\bm{3.10}(2.65)$ & $4.35(3.72)$ & $4.12(3.81)$  & $\bm{2.93}(2.51)$ & $4.47(3.84)$ & $3.51(3.18)$  & $\bm{2.50}(2.16)$ & $4.40(3.82)$ & $2.85(2.47)$  & $\bm{2.21}(1.80)$ & $4.33(3.63)$ & $2.30(1.80)$ \\
 & $D_{1} (10^{-1})$ & $\bm{1.50}(1.32)$ & $2.11(1.84)$ & $2.05(1.93)$  & $\bm{1.39}(1.21)$ & $2.15(1.89)$ & $1.72(1.57)$  & $\bm{1.19}(1.03)$ & $2.13(1.87)$ & $1.38(1.24)$  & $\bm{1.04}(0.87)$ & $2.08(1.83)$ & $1.09(0.88)$ \\
 & $D_{ks} (10^{-2})$ & $\bm{4.04}(3.61)$ & $5.51(4.86)$ & $6.33(5.90)$  & $\bm{3.74}(3.30)$ & $5.60(4.92)$ & $5.16(4.77)$  & $\bm{3.18}(2.78)$ & $5.51(4.83)$ & $4.04(3.64)$  & $\bm{2.88}(2.48)$ & $5.41(4.75)$ & $3.08(2.59)$ \\
 & $D_{cm} (10^{-3})$ & $\bm{0.82}(0.46)$ & $1.82(0.92)$ & $1.83(1.26)$  & $\bm{0.72}(0.37)$ & $1.87(0.95)$ & $1.19(0.83)$  & $\bm{0.52}(0.27)$ & $1.82(0.95)$ & $0.73(0.48)$  & $\bm{0.46}(0.22)$ & $1.76(0.91)$ & $0.50(0.24)$ \\
 & $D_{m} (10^{-5})$ & $1.54(1.23)$ & $\bm{1.11}(0.92)$ & $5.42(4.25)$  & $1.34(1.09)$ & $\bm{1.07}(0.87)$ & $4.08(3.20)$  & $1.14(0.96)$ & $\bm{1.02}(0.85)$ & $2.79(2.21)$  & $1.01(0.84)$ & $\bm{0.99}(0.83)$ & $1.48(1.20)$ \\
 & $D_{v} (10^{-7})$ & $1.31(1.06)$ & $2.10(1.82)$ & $\bm{1.27}(0.87)$  & $1.17(0.92)$ & $2.10(1.86)$ & $\bm{1.08}(0.77)$  & $1.02(0.78)$ & $2.13(1.95)$ & $\bm{0.91}(0.61)$  & $0.79(0.56)$ & $2.06(1.78)$ & $\bm{0.77}(0.50)$ \\
\\
$ T = 200 $ & $D_{2}$ & $\bm{3.13}(2.69)$ & $4.53(3.87)$ & $4.11(3.76)$  & $\bm{2.75}(2.38)$ & $4.50(3.80)$ & $3.42(3.11)$  & $\bm{2.45}(2.04)$ & $4.37(3.72)$ & $2.86(2.46)$  & $\bm{2.02}(1.56)$ & $4.20(3.61)$ & $2.17(1.68)$ \\
 & $D_{1} (10^{-1})$ & $\bm{1.50}(1.31)$ & $2.19(1.87)$ & $2.03(1.88)$  & $\bm{1.31}(1.16)$ & $2.17(1.86)$ & $1.68(1.54)$  & $\bm{1.16}(0.99)$ & $2.11(1.83)$ & $1.38(1.22)$  & $\bm{0.95}(0.76)$ & $2.03(1.81)$ & $1.04(0.83)$ \\
 & $D_{ks} (10^{-2})$ & $\bm{3.95}(3.47)$ & $5.67(4.89)$ & $6.31(5.82)$  & $\bm{3.47}(3.11)$ & $5.63(4.84)$ & $5.04(4.63)$  & $\bm{3.11}(2.66)$ & $5.48(4.76)$ & $4.04(3.63)$  & $\bm{2.65}(2.17)$ & $5.27(4.70)$ & $2.93(2.47)$ \\
 & $D_{cm} (10^{-3})$ & $\bm{0.82}(0.42)$ & $1.95(0.94)$ & $1.83(1.24)$  & $\bm{0.61}(0.34)$ & $1.90(0.95)$ & $1.13(0.76)$  & $\bm{0.51}(0.25)$ & $1.82(0.93)$ & $0.74(0.47)$  & $\bm{0.40}(0.17)$ & $1.66(0.88)$ & $0.45(0.22)$ \\
 & $D_{m} (10^{-5})$ & $1.26(1.03)$ & $\bm{1.02}(0.85)$ & $5.58(4.45)$  & $1.13(0.94)$ & $\bm{1.02}(0.86)$ & $3.96(3.14)$  & $1.03(0.89)$ & $\bm{0.97}(0.83)$ & $2.74(2.26)$  & $\bm{0.96}(0.78)$ & $0.96(0.79)$ & $1.45(1.20)$ \\
 & $D_{v} (10^{-7})$ & $1.27(1.02)$ & $2.15(1.90)$ & $\bm{1.27}(0.88)$  & $1.11(0.90)$ & $2.15(1.93)$ & $\bm{1.05}(0.72)$  & $1.00(0.73)$ & $2.11(1.89)$ & $\bm{0.89}(0.60)$  & $0.76(0.53)$ & $2.07(1.87)$ & $\bm{0.75}(0.51)$ \\
\\
$ T = 500 $ & $D_{2}$ & $\bm{3.02}(2.65)$ & $4.49(3.83)$ & $4.05(3.78)$  & $\bm{2.77}(2.39)$ & $4.44(3.82)$ & $3.46(3.14)$  & $\bm{2.38}(2.03)$ & $4.36(3.69)$ & $2.83(2.47)$  & $\bm{2.02}(1.59)$ & $4.31(3.70)$ & $2.17(1.73)$ \\
 & $D_{1} (10^{-1})$ & $\bm{1.44}(1.30)$ & $2.17(1.90)$ & $2.00(1.86)$  & $\bm{1.31}(1.17)$ & $2.14(1.86)$ & $1.69(1.55)$  & $\bm{1.13}(0.97)$ & $2.10(1.85)$ & $1.37(1.21)$  & $\bm{0.95}(0.78)$ & $2.08(1.81)$ & $1.03(0.84)$ \\
 & $D_{ks} (10^{-2})$ & $\bm{3.77}(3.38)$ & $5.62(4.95)$ & $6.27(5.83)$  & $\bm{3.46}(3.07)$ & $5.56(4.85)$ & $5.11(4.73)$  & $\bm{3.03}(2.59)$ & $5.47(4.77)$ & $3.98(3.62)$  & $\bm{2.65}(2.28)$ & $5.37(4.71)$ & $2.94(2.55)$ \\
 & $D_{cm} (10^{-3})$ & $\bm{0.74}(0.40)$ & $1.91(0.97)$ & $1.80(1.27)$  & $\bm{0.63}(0.34)$ & $1.87(0.92)$ & $1.16(0.81)$  & $\bm{0.48}(0.24)$ & $1.80(0.91)$ & $0.71(0.46)$  & $\bm{0.39}(0.19)$ & $1.75(0.89)$ & $0.44(0.23)$ \\
 & $D_{m} (10^{-5})$ & $1.11(0.95)$ & $\bm{1.01}(0.85)$ & $5.58(4.49)$  & $1.05(0.90)$ & $\bm{0.99}(0.85)$ & $4.07(3.26)$  & $1.01(0.85)$ & $\bm{1.00}(0.84)$ & $2.64(2.11)$  & $0.96(0.80)$ & $\bm{0.96}(0.78)$ & $1.50(1.21)$ \\
 & $D_{v} (10^{-7})$ & $1.24(1.00)$ & $2.13(1.93)$ & $\bm{1.21}(0.83)$  & $1.12(0.84)$ & $2.13(1.97)$ & $\bm{1.04}(0.71)$  & $0.97(0.72)$ & $2.13(1.98)$ & $\bm{0.89}(0.61)$  & $\bm{0.73}(0.51)$ & $2.08(1.90)$ & $0.73(0.49)$ \\
\\
\hline \hline
\end{tabular}
\caption{Monte Carlo Simulations on Out-of-Sample Forecasts (GBP/USD Exchange Rate)}
\label{tab:simu_forex}
\end{table}
\end{landscape}

\begin{landscape}
\begin{table}
\tiny\centering
\begin{tabular}{ll|lll|lll|lll|lll}
\hline \hline
 & & $ N = 100 $  & & & $ N = 200 $  & & & $ N = 500 $  & & & $ N = 5000 $  & &  \\ \hline
 & & FAR & AVE & LAST& FAR & AVE & LAST& FAR & AVE & LAST& FAR & AVE & LAST\\ \hline
$ T = 50 $ & $D_{2} (10^{-1})$ & $\bm{4.25}(3.97)$ & $4.40(4.27)$ & $5.37(4.82)$  & $\bm{4.15}(3.94)$ & $4.37(4.13)$ & $5.20(4.74)$  & $\bm{3.98}(3.70)$ & $4.21(3.99)$ & $4.99(4.45)$  & $\bm{3.95}(3.70)$ & $4.29(4.12)$ & $4.88(4.44)$ \\
 & $D_{1} (10^{-1})$ & $\bm{2.80}(2.63)$ & $2.90(2.79)$ & $3.66(3.30)$  & $\bm{2.74}(2.56)$ & $2.89(2.74)$ & $3.53(3.21)$  & $\bm{2.63}(2.45)$ & $2.78(2.65)$ & $3.35(2.99)$  & $\bm{2.61}(2.44)$ & $2.85(2.75)$ & $3.24(2.91)$ \\
 & $D_{ks} (10^{-1})$ & $\bm{1.28}(1.19)$ & $1.31(1.25)$ & $1.59(1.40)$  & $\bm{1.26}(1.18)$ & $1.30(1.21)$ & $1.56(1.38)$  & $\bm{1.22}(1.12)$ & $1.25(1.17)$ & $1.52(1.32)$  & $\bm{1.22}(1.12)$ & $1.28(1.23)$ & $1.50(1.32)$ \\
 & $D_{cm} (10^{-2})$ & $\bm{0.94}(0.60)$ & $0.95(0.66)$ & $1.61(0.83)$  & $0.97(0.62)$ & $\bm{0.96}(0.64)$ & $1.62(0.83)$  & $0.92(0.57)$ & $\bm{0.91}(0.59)$ & $1.58(0.79)$  & $\bm{0.96}(0.58)$ & $0.99(0.65)$ & $1.56(0.78)$ \\
 & $D_{m} (10^{-2})$ & $\bm{2.83}(2.62)$ & $2.83(2.58)$ & $3.63(3.18)$  & $2.86(2.64)$ & $\bm{2.81}(2.62)$ & $3.60(3.10)$  & $2.78(2.50)$ & $\bm{2.73}(2.50)$ & $3.51(2.98)$  & $\bm{2.82}(2.51)$ & $2.83(2.57)$ & $3.45(2.99)$ \\
 & $D_{v} (10^{-3})$ & $\bm{4.71}(4.07)$ & $5.33(4.75)$ & $5.40(4.14)$  & $\bm{4.21}(3.68)$ & $5.32(4.78)$ & $4.67(3.72)$  & $\bm{3.93}(3.25)$ & $5.12(4.56)$ & $4.27(3.32)$  & $\bm{3.63}(2.97)$ & $5.12(4.54)$ & $4.02(3.23)$ \\
\\
$ T = 100 $ & $D_{2} (10^{-1})$ & $\bm{4.16}(3.91)$ & $4.40(4.15)$ & $5.37(4.91)$  & $\bm{4.08}(3.81)$ & $4.37(4.15)$ & $5.10(4.58)$  & $\bm{4.01}(3.79)$ & $4.37(4.25)$ & $5.07(4.56)$  & $\bm{3.89}(3.68)$ & $4.25(4.12)$ & $4.89(4.42)$ \\
 & $D_{1} (10^{-1})$ & $\bm{2.74}(2.57)$ & $2.90(2.78)$ & $3.67(3.34)$  & $\bm{2.68}(2.49)$ & $2.89(2.75)$ & $3.45(3.09)$  & $\bm{2.64}(2.52)$ & $2.89(2.80)$ & $3.40(3.01)$  & $\bm{2.56}(2.43)$ & $2.82(2.70)$ & $3.24(2.91)$ \\
 & $D_{ks} (10^{-1})$ & $\bm{1.26}(1.18)$ & $1.30(1.22)$ & $1.60(1.41)$  & $\bm{1.24}(1.14)$ & $1.30(1.22)$ & $1.52(1.29)$  & $\bm{1.24}(1.18)$ & $1.31(1.24)$ & $1.55(1.36)$  & $\bm{1.20}(1.11)$ & $1.27(1.18)$ & $1.50(1.33)$ \\
 & $D_{cm} (10^{-2})$ & $\bm{0.93}(0.60)$ & $0.95(0.64)$ & $1.67(0.83)$  & $\bm{0.92}(0.56)$ & $0.96(0.63)$ & $1.52(0.68)$  & $\bm{0.94}(0.63)$ & $0.98(0.69)$ & $1.61(0.83)$  & $\bm{0.91}(0.56)$ & $0.95(0.62)$ & $1.56(0.82)$ \\
 & $D_{m} (10^{-2})$ & $\bm{2.81}(2.58)$ & $2.82(2.62)$ & $3.69(3.09)$  & $\bm{2.78}(2.49)$ & $2.82(2.56)$ & $3.48(2.86)$  & $\bm{2.85}(2.67)$ & $2.87(2.70)$ & $3.59(3.21)$  & $\bm{2.75}(2.52)$ & $2.78(2.55)$ & $3.45(2.97)$ \\
 & $D_{v} (10^{-3})$ & $\bm{4.46}(3.99)$ & $5.42(4.97)$ & $5.13(3.90)$  & $\bm{4.21}(3.51)$ & $5.27(4.59)$ & $4.84(3.72)$  & $\bm{3.80}(3.11)$ & $5.16(4.53)$ & $4.28(3.45)$  & $\bm{3.55}(2.80)$ & $5.00(4.41)$ & $4.03(3.21)$ \\
\\
$ T = 200 $ & $D_{2} (10^{-1})$ & $\bm{4.13}(3.87)$ & $4.39(4.23)$ & $5.30(4.73)$  & $\bm{4.03}(3.72)$ & $4.35(4.20)$ & $5.14(4.55)$  & $\bm{4.10}(3.85)$ & $4.45(4.26)$ & $5.15(4.61)$  & $\bm{3.83}(3.60)$ & $4.23(4.10)$ & $4.93(4.40)$ \\
 & $D_{1} (10^{-1})$ & $\bm{2.70}(2.54)$ & $2.89(2.78)$ & $3.61(3.23)$  & $\bm{2.64}(2.45)$ & $2.87(2.79)$ & $3.46(3.09)$  & $\bm{2.68}(2.54)$ & $2.94(2.86)$ & $3.44(3.06)$  & $\bm{2.53}(2.35)$ & $2.81(2.73)$ & $3.26(2.87)$ \\
 & $D_{ks} (10^{-1})$ & $\bm{1.25}(1.17)$ & $1.30(1.22)$ & $1.56(1.35)$  & $\bm{1.23}(1.13)$ & $1.29(1.21)$ & $1.53(1.34)$  & $\bm{1.26}(1.20)$ & $1.33(1.29)$ & $1.57(1.39)$  & $\bm{1.19}(1.09)$ & $1.27(1.18)$ & $1.51(1.33)$ \\
 & $D_{cm} (10^{-2})$ & $\bm{0.90}(0.58)$ & $0.95(0.63)$ & $1.56(0.79)$  & $\bm{0.90}(0.56)$ & $0.95(0.62)$ & $1.56(0.80)$  & $\bm{0.97}(0.66)$ & $1.02(0.74)$ & $1.65(0.88)$  & $\bm{0.89}(0.56)$ & $0.95(0.64)$ & $1.57(0.80)$ \\
 & $D_{m} (10^{-2})$ & $\bm{2.77}(2.55)$ & $2.81(2.54)$ & $3.54(3.08)$  & $\bm{2.76}(2.48)$ & $2.78(2.57)$ & $3.53(3.07)$  & $\bm{2.89}(2.74)$ & $2.93(2.81)$ & $3.64(3.24)$  & $\bm{2.75}(2.50)$ & $2.80(2.58)$ & $3.49(3.03)$ \\
 & $D_{v} (10^{-3})$ & $\bm{4.50}(3.95)$ & $5.37(4.77)$ & $5.30(4.05)$  & $\bm{4.08}(3.41)$ & $5.27(4.70)$ & $4.80(3.66)$  & $\bm{3.92}(3.22)$ & $5.27(4.59)$ & $4.40(3.36)$  & $\bm{3.44}(2.72)$ & $4.98(4.38)$ & $4.00(3.21)$ \\
\\
$ T = 500 $ & $D_{2} (10^{-1})$ & $\bm{4.17}(4.00)$ & $4.46(4.32)$ & $5.44(4.96)$  & $\bm{4.15}(3.95)$ & $4.49(4.31)$ & $5.29(4.80)$  & $\bm{3.90}(3.71)$ & $4.28(4.11)$ & $4.96(4.46)$  & $\bm{3.81}(3.50)$ & $4.20(3.95)$ & $4.85(4.27)$ \\
 & $D_{1} (10^{-1})$ & $\bm{2.72}(2.62)$ & $2.94(2.88)$ & $3.71(3.39)$  & $\bm{2.71}(2.58)$ & $2.95(2.88)$ & $3.58(3.21)$  & $\bm{2.57}(2.43)$ & $2.84(2.73)$ & $3.33(2.99)$  & $\bm{2.50}(2.29)$ & $2.78(2.62)$ & $3.20(2.82)$ \\
 & $D_{ks} (10^{-1})$ & $\bm{1.27}(1.20)$ & $1.33(1.27)$ & $1.61(1.41)$  & $\bm{1.27}(1.20)$ & $1.34(1.28)$ & $1.60(1.42)$  & $\bm{1.20}(1.12)$ & $1.28(1.21)$ & $1.51(1.32)$  & $\bm{1.17}(1.06)$ & $1.25(1.17)$ & $1.49(1.28)$ \\
 & $D_{cm} (10^{-2})$ & $\bm{0.92}(0.63)$ & $0.97(0.68)$ & $1.67(0.89)$  & $\bm{0.95}(0.66)$ & $1.01(0.74)$ & $1.67(0.88)$  & $\bm{0.90}(0.57)$ & $0.96(0.64)$ & $1.54(0.75)$  & $\bm{0.87}(0.52)$ & $0.94(0.62)$ & $1.53(0.71)$ \\
 & $D_{m} (10^{-2})$ & $\bm{2.84}(2.64)$ & $2.87(2.71)$ & $3.68(3.25)$  & $\bm{2.87}(2.75)$ & $2.92(2.78)$ & $3.69(3.32)$  & $\bm{2.76}(2.56)$ & $2.82(2.54)$ & $3.48(3.01)$  & $\bm{2.71}(2.53)$ & $2.77(2.58)$ & $3.42(2.86)$ \\
 & $D_{v} (10^{-3})$ & $\bm{4.47}(3.81)$ & $5.50(4.91)$ & $5.24(3.98)$  & $\bm{4.24}(3.46)$ & $5.42(4.85)$ & $4.88(3.73)$  & $\bm{3.81}(3.11)$ & $5.23(4.65)$ & $4.33(3.46)$  & $\bm{3.51}(2.82)$ & $4.90(4.25)$ & $3.99(3.04)$ \\
\\
\hline \hline
\end{tabular}
\caption{Monte Carlo Simulations on Out-of-Sample Forecasts (NYSE Stock Returns)}
\label{tab:simu_nyse}
\end{table}
\end{landscape}

\newpage
\appendix

\section*{Appendix: Mathematical Proofs}
\begin{proof}[Proof of Lemma \ref{lemma:1}]
  Let $\tilde f = (1/T)\sum_{t=1}^T f_t$. We first show the $L^2$ consistency. Notice that
  \begin{equation*}
    \norm{\bar f- \mathbb{E} f}^2 = \norm{\bar f- \tilde f + \tilde f - \mathbb{E} f}^2 \leq 2\left(\norm{\bar f- \tilde f}^2 + \norm{\tilde f- \mathbb{E} f}^2\right).
  \end{equation*}
By Assumption \ref{assump_2}\ref{ass2_c}, we have that
\begin{equation*}
  \mathbb{E}\norm{\bar f- \tilde f}^2 = \mathbb{E}\norm{\frac{1}{T}\sum_{i=1}^T\left(\hat f_t - f_t\right)}^2 \leq \mathbb{E} \frac{1}{T} \sum_{i=1}^T \norm{\Delta_t}^2 \leq \sup_t \mathbb{E} \norm{\Delta_t}^2 = O\left(N^{-r}\right).
\end{equation*}
If $N\geq c T^{1/r}$ for some constant $c>0$, then we have that $\mathbb{E}\norm{\bar f - \tilde f}^2 = O(T^{-1})$. By Theorem 3.7 in \cite{bosq-00}, we have that $\mathbb{E}\norm{\tilde f-\mathbb{E} f}^2 = O(T^{-1})$. Therefore, $\mathbb{E}\norm{\bar f - \mathbb{E} f}^2 = O(T^{-1})$.

Next we show the a.s. consistency. Notice that
\begin{equation*}
  \norm{\bar f - \mathbb{E} f} \leq \norm{\bar f - \tilde f} + \norm{\tilde f - \mathbb{E} f}.
\end{equation*}
By Chebyshev's inequality, for any $\epsilon>0$,
\begin{equation*}
  \mathbb{P}\left\{\left(\frac{T}{\ln T}\right)^{1/2} \norm{\bar f- \tilde f}\geq \epsilon\right\} \leq \frac{T}{\epsilon^2\ln T}\mathbb{E}\norm{\bar f - \tilde f}^2 = \frac{T}{\epsilon^2 \ln T} O(N^{-r}).
\end{equation*}
If $N>cT^{2/r}\ln^s T$ for some constants $c>0$ and $s>0$, then
\begin{equation*}
  \mathbb{P}\left\{\left(\frac{T}{\ln T}\right)^{1/2} \norm{\bar f- \tilde f}\geq \epsilon\right\} = O\left(\frac{1}{T\ln^\delta T}\right)
\end{equation*}
where $\delta = 1+sr>1$. Since $\sum_{T=2}^\infty 1/T\ln^\delta T$ converges for any $\delta>1$, by the Borel-Cantelli lemma, we have that
\begin{equation*}
  \mathbb{P}\left\{ \limsup_{T\to \infty} \left(\frac{T}{\ln T}\right)^{1/2} \norm{\bar f- \tilde f}\geq \epsilon\right\} = 0.
\end{equation*}
This implies that $\norm{\bar f-\tilde f} = o(T^{-1/2}\ln^{1/2} T)$ a.s.. By Corollary 3.2 in \cite{bosq-00}, we have that $\norm{\tilde f - \mathbb{E} f} = O(T^{-1/2} \ln^{1/2} T)$ a.s.. Therefore, we conclude that $\norm{\bar f - \mathbb{E} f} = O(T^{-1/2} \ln^{1/2} T)$ a.s..
\end{proof}


\begin{proof}[Proof of Theorem \ref{thm:2}]
  We first prove the $L^2$ consistency of $\hat Q$. By Assumption \ref{assump_2}\ref{ass2_b}, we have that $\norm{\mathbb{E} f}\leq \mathbb{E}\norm{f_t} \leq M$. Then for all $t$,
  \begin{equation}
    \label{eq:w_bound}
    \norm{w_t} = \norm{f_t-\mathbb{E} f} \leq \norm{f_t} + \norm{\mathbb{E} f} \leq 2M \quad \text{a.s.}.
  \end{equation}
  Similarly, we have that for all $t$,
  \begin{equation*}
    \norm{\hat f_t} = \norm{f_t + \Delta_t} \leq 2M \quad \text{a.s.}
  \end{equation*}
  and therefore
  \begin{equation}
    \label{eq:w_hat_bound}
    \norm{\hat w_t} = \norm{\hat f_t - \bar f} = \norm{\hat f_t - \frac{1}{T}\sum_{i=1}^T \hat f_i} \leq 4M \quad \text{a.s.}.
  \end{equation}
 By Assumption \ref{assump_2}\ref{ass2_c} and Lemma \ref{lemma:1},
  \begin{equation}
    \label{eq:dw_t_MSE}
    \mathbb{E} \norm{\hat w_t - w_t}^2 = \norm{\hat f_t - \bar f - f_t + \mathbb{E} f}^2 \leq 2\left(\mathbb{E}\norm{\hat f_t - f_t}^2 + \mathbb{E} \norm{\bar f- \mathbb{E} f}^2\right) = O(T^{-1}).
  \end{equation}

  Let
  \begin{equation*}
    \tilde Q = \frac{1}{T}\sum_{t=1}^T w_t\otimes w_t.
  \end{equation*}
  then
  \begin{align*}
    \mathbb{E}\norm{\hat Q - \tilde Q}^2
    & = \mathbb{E} \norm{\frac{1}{T}\sum_{t=1}^T (\hat w_t\otimes \hat w_t - w_t\otimes w_t)}^2 \\
    & = \frac{1}{T^2}\mathbb{E} \norm{\sum_{t=1}^T \left[\hat w_t\otimes (\hat w_t - w_t) + (\hat w_t - w_t)\otimes w_t \right]}^2 \\
    & \leq \frac{2}{T} \mathbb{E}\left(\sum_{t=1}^T\norm{\hat w_t\otimes (\hat w_t-w_t)}^2 + \sum_{t=1}^T \norm{(\hat w_t-w_t) \otimes w_t}^2\right) \\
    & \leq \frac{2}{T}\mathbb{E} \left(\sum_{t=1}^T \left(\norm{\hat w_t}^2 + \norm{w_t}^2\right)\norm{\hat w_t-w_t}^2\right) \\
    & \leq \frac{40M^2}{T}\sum_{t=1}^T\mathbb{E}\norm{\hat w_t-w_t}^2 = O(T^{-1}).
  \end{align*}
  By Theorem 4.1 in \cite{bosq-00}, we have that
  \begin{equation*}
    \mathbb{E}\norm{\tilde Q - Q}^2 = O\left(T^{-1}\right).
  \end{equation*}
  Therefore,
  \begin{equation*}
    \mathbb{E}\norm{\hat Q - Q}^2 \leq 2\left(\mathbb{E}\norm{\hat Q - \tilde Q}^2 + \mathbb{E}\norm{\tilde Q - Q}^2\right) = O\left(T^{-1}\right)
  \end{equation*}
  as $T\to \infty$.

  Next we show the a.s. consistency of $\hat Q$. First note that
  \begin{align*}
    \norm{\hat Q - \tilde Q}
    & = \norm{\frac{1}{T}\sum_{t=1}^T (\hat w_t\otimes \hat w_t - w_t\otimes w_t)} \\
    & = \norm{\frac{1}{T}\sum_{t=1}^T [\hat w_t\otimes (\hat w_t- w_t) + (\hat w_t - w_t)\otimes w_t]} \\
    & \leq \frac{1}{T} \sum_{t=1}^T \left(\norm{\hat w_t} + \norm{w_t}\right) \norm{\hat w_t - w_t} \\
    & \leq \frac{6M}{T}\sum_{t=1}^T \norm{\hat w_t - w_t} \quad \text{a.s.}.
  \end{align*}
Since
  \begin{equation*}
    \mathbb{E}\left(\frac{1}{T}\sum_{t=1}^T \norm{\hat f_t - f_t}\right)^2 \leq \mathbb{E} \frac{1}{T^2}\cdot T\sum_{t=1}^T \norm{\Delta_t}^2 \leq \sup_t \mathbb{E}\norm{\Delta_t}^2 = O(N^{-r}),
  \end{equation*}
  by Chebyshev's inequality, we have that for any $\epsilon>0$,
\begin{equation*}
  \mathbb{P}\left\{\left(\frac{T}{\ln T}\right)^{1/2} \frac{1}{T}\sum_{t=1}^T \norm{\hat f_t - f_t}\geq \epsilon\right\} \leq \frac{T}{\epsilon^2\ln T}\mathbb{E}\left(\frac{1}{T}\sum_{t=1}^T \norm{\hat f_t - f_t}\right)^2 = \frac{T}{\epsilon^2 \ln T} O(N^{-r}).
\end{equation*}
If $N>cT^{2/r}\ln^s T$ for some constants $c>0$ and $s>0$, by the Borel-Cantelli lemma, we have that
\begin{equation*}
  \mathbb{P}\left\{\limsup_{T\to \infty} \left(\frac{T}{\ln T} \right)^{1/2} \frac{1}{T}\sum_{t=1}^T \norm{\hat f_t - f_t}\geq \epsilon\right\} = 0.
\end{equation*}
This implies that
\begin{equation*}
  \frac{1}{T}\sum_{t=1}^T \norm{\hat f_t - f_t} = o\left(T^{-1/2}\ln^{1/2} T\right) \quad \text{a.s.}.
\end{equation*}
By Lemma \ref{lemma:1}, $\norm{\bar f - \mathbb{E} f} = O\left(T^{-1/2}\ln^{1/2} T\right)$ a.s.. Then
\begin{gather}
\label{eq:w_dev_sum}
\begin{aligned}
  \frac{1}{T}\sum_{t=1}^T\norm{\hat w_t - w_t}
  & = \frac{1}{T}\sum_{t=1}^T\norm{\hat f_t - \bar f -f_t + \mathbb{E} f} \\
  & \leq \frac{1}{T}\sum_{t=1}^T \norm{\hat f_t - f_t} + \norm{\bar f - \mathbb{E} f} \\
  & = O\left(T^{-1/2}\ln^{1/2} T\right)\quad \text{a.s.}.
\end{aligned}
\end{gather}
By Corollary 4.1 in \cite{bosq-00}, $\norm{\tilde Q - Q} = O\left(T^{-1/2}\ln^{1/2} T\right)$ a.s.. Therefore
\begin{equation*}
    \norm{\hat Q - Q} \leq \norm{\hat Q - \tilde Q} + \norm{\tilde Q - Q} = O\left(T^{-1/2}\ln^{1/2} T\right) \quad \text{a.s.}.
  \end{equation*}

The proof for consistency of $\hat P$ is similar and therefore is omitted. The proof makes uses of Theorem 4.7 and 4.8 in \cite{bosq-00}.
\end{proof}


\begin{proof}[Proof of Lemma \ref{lemma:3}]
  See Lemma 4.2 and Lemma 4.3 in \cite{bosq-00}.
\end{proof}


\begin{proof}[Proof of Theorem \ref{thm:4}]
  It follows immediately from Theorem \ref{thm:2} and Lemma \ref{lemma:3}.
\end{proof}


\begin{proof}[Proof of Theorem \ref{thm:5}]
  Notice that
  \begin{align*}
    \norm{\hat A_K - A} & =  \norm{\hat P\sum_{k=1}^K \hat \lambda_k^{-1}(\hat v_k \otimes \hat v_k) - A} \\
    & \leq \norm{D_1} + \norm{D_2} + \norm{D_3} + \norm{D_4}
  \end{align*}
  where
  \begin{equation*}
    D_1 = \hat P\sum_{k=1}^K \hat \lambda_k^{-1}(\hat v_k \otimes \hat v_k) - P\sum_{k=1}^K \hat \lambda_k^{-1}(\hat v_k \otimes \hat v_k),
  \end{equation*}
  \begin{equation*}
    D_2  = P\sum_{k=1}^K \hat \lambda_k^{-1}(\hat v_k \otimes \hat v_k) - P\sum_{k=1}^K \lambda_k^{-1}(\hat v_k \otimes \hat v_k),
  \end{equation*}
  \begin{equation*}
    D_3 = P\sum_{k=1}^K \lambda_k^{-1}(\hat v_k \otimes \hat v_k) - P\sum_{k=1}^K \lambda_k^{-1}(v_k \otimes v_k),
  \end{equation*}
  and
  \begin{equation*}
    D_4 = P\sum_{k=1}^K \lambda_k^{-1}(v_k \otimes v_k) - A.
  \end{equation*}

Since $\hat \lambda_1 \geq \hat \lambda_2 \geq \cdots >0$, we have that
\begin{align*}
  \norm{D_1}
  & \leq \norm{\hat P - P}\norm{\sum_{i=1}^K \hat \lambda_k^{-1}(\hat v_k \otimes \hat v_k)} \\
  & = \hat \lambda_K^{-1} \norm{\hat P - P} \\
  & \leq \lambda_K^{-1} \norm{\hat P - P} + \abs{\hat \lambda_K^{-1} - \lambda_K^{-1}}\norm{\hat P- P}.
\end{align*}

Note that
\begin{equation*}
  \sum_{k = 1}^K \tau_k \geq 2\sqrt{2}(\lambda_K - \lambda_{K+1})^{-1} \geq 2\sqrt{2}\lambda_K^{-1},
\end{equation*}
then by assumption we have that
\begin{equation*}
  \frac{\ln T \left(\sum_{k=1}^K \tau_k\right)^2}{T\lambda_K^2} \geq \frac{8\ln T}{T\lambda_K^4} \to 0
\end{equation*}
as $T\to \infty$. This implies that
\begin{equation}
  \label{eq:lambda_K}
  \lambda_K^{-1} = o\left(T^{1/4}\ln^{-1/4} T\right).
\end{equation}
 By Theorem 2, we have that $
\norm{\hat P - P} = O\left(T^{-1/2}\ln^{1/2} T\right)$ a.s.. Then
\begin{equation}
  \label{eq:D1_part1}
  \lambda_K^{-1}\norm{\hat P - P} \to 0 \quad \text{ a.s.}
\end{equation}
as $T\to \infty$.

Next we notice that
\begin{equation}
  \label{eq:hat_lambda_close_to_lambda}
  \mathbb{P}\left(\liminf_{k\to \infty} \left\{\hat \lambda_k \geq \frac{\lambda_k}{2}\right\}\right) = 1,
\end{equation}
for if otherwise, then with positive probability, $\hat \lambda_k < \lambda_k/2$ infinitely often. This implies that with positive probability, $\abs{\hat \lambda_k - \lambda_k} > \lambda_k/2$ infinitely often. Then by \eqref{eq:lambda_K}, we have that with positive probability, $\limsup_{T\to \infty} T^{1/4}\ln^{-1/4} T \abs{\hat \lambda_k - \lambda_k} = \infty$. This contradicts with the conclusion in Theorem \ref{thm:4}. Therefore, \eqref{eq:hat_lambda_close_to_lambda} holds. That is, almost surely, $\hat \lambda_K \geq \lambda_K/2$ for $K$ large enough.

Now for such $K$,
\begin{align*}
  \abs{\hat \lambda_K^{-1} - \lambda_K^{-1}}\norm{\hat P- P}
  & = \frac{\abs{\hat \lambda_K - \lambda_K}}{\hat \lambda_K \lambda_K}\norm{\hat P- P} \\
  & \leq 2\frac{\abs{\hat \lambda_K - \lambda_K}}{ \lambda_K^2}\norm{\hat P- P} \\
  & \leq \frac{2\norm{\hat Q - Q}\norm{\hat P - P}}{\lambda_K^2}.
\end{align*}
Then by Theorem \ref{thm:2} and equation \eqref{eq:lambda_K},
\begin{equation}
  \label{eq:D1_part2}
  \abs{\hat \lambda_K^{-1} - \lambda_K^{-1}}\norm{\hat P- P} \to 0 \quad \text{ a.s.}.
\end{equation}
Equations \eqref{eq:D1_part1} and \eqref{eq:D1_part2} together imply that
\begin{equation*}
\norm{D_1} \to 0 \quad \text{a.s.}
\end{equation*}
as $T\to \infty$.

For $K$ large enough,
\begin{align*}
  \norm{D_2}
  & \leq \norm{P}\norm{\sum_{k=1}^K \left(\hat \lambda_k^{-1} - \lambda_k^{-1}\right)(\hat v_k \otimes \hat v_k)} \\
  & = \max_{1\leq k\leq K} \abs{\hat \lambda_k^{-1} - \lambda_k^{-1}}\norm{P} \\
  & \leq \frac{\sup_{k\geq 1}\abs{\hat \lambda_k - \lambda_k}\norm{P}}{\hat \lambda_K \lambda_K} \\
  & \leq \frac{2\norm{\hat Q - Q}\norm{P}}{ \lambda_K^2}.
\end{align*}
By Theorem \ref{thm:2} and equation \eqref{eq:lambda_K},
\begin{equation*}
  \norm{D_2} \to 0 \quad \text{a.s.}
\end{equation*}
as $T\to \infty$.

By Lemma 3,
\begin{align*}
  \norm{D_3}
  & \leq \norm{P}\sum_{k=1}^K\lambda_k^{-1}\norm{\hat v_k\otimes \hat v_k - v_k'\otimes v_k'} \\
  & = \norm{P}\sum_{k=1}^K\lambda_k^{-1}\norm{\hat v_k\otimes (\hat v_k - v_k') + (\hat v_k - v_k') \otimes v_k'} \\
  & \leq 2\norm{P} \lambda_K^{-1} \sum_{k=1}^K \norm{\hat v_k - v_k'} \\
  & \leq 2\lambda_K^{-1}\norm{P}\norm{\hat Q - Q}\sum_{k=1}^K \tau_k.
\end{align*}
Theorem 2 and the assumption that $\ln T \left(\sum_{i=1}^K \tau_k\right)^2/T\lambda_K^2 \to 0$ then imply that
\begin{equation*}
  \norm{D_3} \to 0 \quad \text{a.s.}
\end{equation*}
as $T\to \infty$.

In the end, notice that $Av_k = P\lambda_k^{-1}v_k$, then
\begin{align*}
  \norm{D_4} & = \sup_{\norm{x}\leq 1} \abs{P\sum_{k=1}^K \lambda_k^{-1}(v_k\otimes v_k)(x)-A(x)} \\
  & = \sup_{\{\alpha_k:\sum_{k=1}^\infty \alpha_k^2 \leq 1\}}\abs{P\sum_{k=1}^K \lambda_k^{-1}(v_k\otimes v_k)\left(\sum_{k=1}^\infty \alpha_k v_k\right) - A\left(\sum_{k=1}^\infty \alpha_k v_k\right)} \\
  & = \sup_{\{\alpha_k:\sum_{k=1}^\infty \alpha_k^2 \leq 1\}}\abs{\sum_{k=1}^K \alpha_k P \lambda_k^{-1}v_k - \sum_{k=1}^\infty \alpha_k Av_k} \\
  & = \sup_{\{\alpha_k:\sum_{k=1}^\infty \alpha_k^2 \leq 1\}}\abs{\sum_{k=1}^K \alpha_k Av_k - \sum_{k=1}^\infty \alpha_k Av_k} \\
  & = \sup_{\{\alpha_k:\sum_{k=1}^\infty \alpha_k^2 \leq 1\}}\abs{\sum_{k>K}^\infty \alpha_k Av_k} \\
  & \leq \sup_{\{\alpha_k:\sum_{k=1}^\infty \alpha_k^2 \leq 1\}} \left(\sum_{k>K}^\infty \alpha_k^2\right)^{1/2} \left(\sum_{k>K}^\infty \abs{Av_k}^2\right)^{1/2} \\
  & \leq \left(\sum_{k>K}^\infty \abs{Av_k}^2\right)^{1/2}.
\end{align*}

Since $A$ is Hilbert-Schmidt, $\sum_{k=1}^\infty \abs{Av_k}^2 < \infty$. This implies that $\sum_{k>K}^\infty \abs{Av_k}^2 \to 0$ as $T\to \infty$. Then
\begin{equation*}
  \norm{D_4} \to 0
\end{equation*}
as $T\to \infty$. Then
\begin{equation*}
  \norm{\hat A_K - A} \to 0 \quad \text{a.s.}
\end{equation*}
as $T\to \infty$.
\end{proof}


\begin{proof}[Proof of Corollary \ref{coro:6}]
By Theorem \ref{thm:5}, we have that $\norm{\hat A_k - A} \to 0$ a.s. as $T\to \infty$. This implies that
\begin{equation*}
  \norm{\hat A_K} = O(1) \quad \text{a.s.}.
\end{equation*}
In the proof of Theorem \ref{thm:2}, we showed that $\norm{w_t}\leq 2M$ a.s. and that $\norm{\hat w_t}\leq 4M$ a.s.. Consequently, we have that
\begin{equation*}
  \norm{\varepsilon_t} = \norm{w_t - Aw_{t-1}} \leq \norm{w_t} + \norm{A}\norm{w_{t-1}} = O(1) \quad \text{a.s.}.
\end{equation*}
and that
\begin{equation*}
  \norm{\hat \varepsilon_t} = \norm{\hat w_t - \hat A_K\hat w_{t-1}} \leq \norm{\hat w_t} + \norm{\hat A_K}\norm{\hat w_{t-1}} = O(1) \quad \text{a.s.}.
\end{equation*}
Also,
\begin{align*}
  \frac{1}{T}\sum_{t=1}^T \norm{\hat \varepsilon_t - \varepsilon_t}
  & = \frac{1}{T}\sum_{t=1}^T \norm{(\hat w_t - w_t) - \hat A_K(\hat w_{t-1} - w_{t-1}) + (A-\hat A_K)w_{t-1}} \\
  & \leq \frac{1}{T}\sum_{t=1}^T\norm{\hat w_t - w_t} + \frac{\norm{\hat A_K}}{T}\sum_{t=1}^T\norm{\hat w_{t-1} - w_{t-1}} + \frac{\norm{A - \hat A_K}}{T}\sum_{t=1}^T \norm{w_{t-1}} \\
  & \to 0 \quad \text{a.s.}
\end{align*}
by \eqref{eq:w_dev_sum} and Theorem \ref{thm:5}. Then,
\begin{align*}
  \norm{\frac{1}{T}\sum_{t=1}^T \hat \varepsilon_t\otimes \hat \varepsilon_t - \frac{1}{T}\sum_{t=1}^T \varepsilon_t\otimes \varepsilon_t}
  & = \norm{\frac{1}{T}\sum_{t=1}^T\left[ \hat \varepsilon_t\otimes (\hat \varepsilon_t - \varepsilon_t) + (\hat \varepsilon_t - \varepsilon_t) \otimes \varepsilon_t\right] } \\
  & \leq \frac{1}{T}\sum_{t=1}^T\left(\norm{\hat \varepsilon_t} + \norm{\varepsilon_t}\right)\norm{\hat \varepsilon_t - \varepsilon_t} \\
  & \to 0 \quad \text{a.s.}.
\end{align*}

By Theorem 2.4 in \cite{bosq-00},
\begin{equation*}
  \norm{\frac{1}{T}\sum_{t=1}^T \varepsilon_t\otimes \varepsilon_t - \mathbb{E}(\varepsilon_t \otimes \varepsilon_t)} \to 0 \quad \text{a.s.}
\end{equation*}
as $T\to \infty$. Then
\begin{align*}
  \norm{\hat \Sigma - \Sigma}
  & \leq \norm{\frac{1}{T}\sum_{t=1}^T \hat \varepsilon_t\otimes \hat \varepsilon_t - \frac{1}{T}\sum_{t=1}^T \varepsilon_t\otimes \varepsilon_t} + \norm{\frac{1}{T}\sum_{t=1}^T \varepsilon_t\otimes \varepsilon_t - \mathbb{E}(\varepsilon_t \otimes \varepsilon_t)} \\
  & \to 0 \quad \text{a.s.}
\end{align*}
as $T\to \infty$.
\end{proof}


\begin{proof}[Proof of Lemma \ref{lemma:7}]
  Write
  \begin{equation*}
    \hat w_t = A\hat w_{t-1} + \varepsilon_t + (\hat w_t - w_t) - A(\hat w_{t-1} - w_{t-1}),
  \end{equation*}
from which we may deduce that
\begin{equation}
  \label{eq:decompose_hat_P}
  \hat P = A\hat Q + S + (R_1+R_2+R_3),
\end{equation}
where
\[
 S = \frac{1}{T}\sum_{t=1}^T \varepsilon_t\otimes w_{t-1},
\]
and
\begin{align*}
 R_1 &= \frac{1}{T}\sum_{t=1}^T\varepsilon_t\otimes (\hat w_{t-1} - w_{t-1}), \\
 R_2 &= \frac{1}{T}\sum_{t=1}^T (\hat w_t - w_t) \otimes \hat w_{t-1}, \\
 R_3 &= -\left[\frac{1}{T}\sum_{t=1}^T A(\hat w_{t-1} - w_{t-1}) \otimes \hat w_{t-1}\right].
\end{align*}
Since
\begin{align*}
  \hat A_K & = \hat P \hat Q_K^+ \\
  & = A\hat Q\hat Q_K^+ + S\hat Q_K^+ + (R_1+R_2+R_3)\hat Q_K^+\\
  & = A\hat \Pi_K + S(\hat Q_K^+ - Q_K^+) + SQ_K^+ + (R_1+R_2+R_3)\hat Q_K^+,
\end{align*}
we have that
\begin{align*}
 \left(\hat A_K-A\hat\Pi_K\right)(w_T) &= SQ_K^+(w_T) + (R_1+R_2+R_3)\hat Q_K^+(w_T)
 + S(\hat Q_K^+ - Q_K^+)(w_T).
\end{align*}

It has been established in \cite{mas-07} that
\begin{equation*}
  \sqrt{T/K}SQ_K^+(w_T) \to_d \mathbb{N}(0, \Sigma).
\end{equation*}
To show that
\begin{equation}
  \label{eq:lemma7_con1}
  \sqrt{T/K}\left(\hat A_K\hat w_T - A\hat \Pi_Kw_T\right) \to_d \mathbb{N}(0, \Sigma),
\end{equation}
it suffices to show that
\begin{equation}
  \label{eq:AN_1}
  \sqrt{T/K}(R_1+R_2+R_3)\hat Q_K^+(w_T) \to_p 0,
\end{equation}
\begin{equation}
  \label{eq:AN_2}
  \sqrt{T/K}S(\hat Q_K^+ - Q_K^+)(w_T) \to_p 0,
\end{equation}
and that
\begin{equation}
  \label{eq:AN_3}
  \sqrt{T/K}\hat A_K(\hat w_T-w_T) \to_p 0.
\end{equation}

To show that \eqref{eq:AN_1} holds, we first note that for $R_1$, we have that
\begin{align*}
  \norm{R_1} & = \norm{\frac{1}{T} \sum_{t=1}^T \varepsilon_t\otimes [\Delta_t - (\bar f - \mathbb{E} f)]} \\
  & \leq \norm{\frac{1}{T} \sum_{t=1}^T \varepsilon_t\otimes \Delta_t } + \norm{\frac{1}{T} \sum_{t=1}^T \varepsilon_t}\norm{ (\bar f - \mathbb{E} f)}.
\end{align*}

Since $\mathbb{E}\norm{\varepsilon_t}^2<\infty$, and since that $\sup_{t\geq 1} \mathbb{E}{\norm{\Delta_t}^2} = O(N^{-r})$ and $N\geq cT^{2/r}\ln^s T$ for some constant $c>0$ and  $s>0$ together imply that $\sup_{t\geq 1}\mathbb{E}{\norm{\Delta_t}^2} = o(T^{-2})$, we have
\begin{align*}
  \mathbb{E}\norm{\frac{1}{T} \sum_{t=1}^T \varepsilon_t\otimes \Delta_t }
  & \leq \frac{1}{T}\sum_{t=1}^T \mathbb{E}\norm{\varepsilon_t}\norm{\Delta_t} \\
  & \leq \frac{1}{T}\sum_{t=1}^T \left(\mathbb{E}\norm{\varepsilon_t}^2\right)^{1/2} \left( \mathbb{E} \norm{\Delta_t}^2\right)^{1/2} \\
  & \leq \left(\mathbb{E} \norm{\varepsilon_t}^2\right)^{1/2} \left(\sup_{t\geq 1} \mathbb{E}\norm{\Delta_t}^2\right)^{1/2} = o(T^{-1}).
\end{align*}
This implies that
\begin{equation*}
  \norm{\frac{1}{T} \sum_{t=1}^T \varepsilon_t\otimes \Delta_t } = o_p(T^{-1}).
\end{equation*}

By Theorem 2.7 in \citet{bosq-00} we have that
\begin{equation*}
  \norm{\frac{1}{T}\sum_{t=1}^T \varepsilon_t}= O_p(T^{-1/2}).
\end{equation*}
Also we have proved in Lemma \ref{lemma:1} that $\norm{\bar f - \mathbb{E} f} = O(T^{-1/2}\ln^{1/2} T)$ a.s.. Then we have that $\norm{R_1} = O_p(T^{-1}\ln^{1/2}T)$.

For $R_2$, write
\begin{equation}
  \label{eq:R_2}
  R_2 =\frac{1}{T}\sum_{t=1}^T (\hat w_t - w_t) \otimes w_{t-1} + \frac{1}{T}\sum_{t=1}^T (\hat w_t - w_t) \otimes (\hat w_{t-1} - w_{t-1}).
\end{equation}
For the first term on the right hand side of \eqref{eq:R_2}, we have that
\begin{align*}
  \norm{\frac{1}{T}\sum_{t=1}^T (\hat w_t - w_t)\otimes w_{t-1}}
  & \leq \norm{\frac{1}{T}\sum_{t=1}^T \Delta_t\otimes w_{t-1}} + \norm{\bar f - \mathbb{E} f}\norm{\frac{1}{T}\sum_{t=1}^Tw_{t-1}}.
\end{align*}
Since $\norm{w_t}^2 \leq 4M^2$ a.s. for all $t$, we have that
\begin{equation}
  \label{eq:sum_delta_w}
\begin{aligned}
  \mathbb{E}\norm{\frac{1}{T}\sum_{t=1}^T \Delta_t\otimes w_{t-1}}
  & \leq \frac{1}{T}\sum_{t=1}^T\mathbb{E}\norm{\Delta_t}\norm{w_{t-1}} \\
  & \leq \frac{1}{T}\sum_{t=1}^T\left(\mathbb{E}\norm{\Delta_t}^2\right)^{1/2} \left(\mathbb{E}\norm{w_{t-1}}^2\right)^{1/2} \\
  & \leq 2M\left(\sup_{k\geq 1} \mathbb{E} \norm{\Delta_t}^2\right)^{1/2} = o(T^{-1}).
\end{aligned}
\end{equation}
This implies that
\begin{equation*}
  \norm{\frac{1}{T}\sum_{t=1}^T \Delta_t\otimes w_{t-1}} = o_p(T^{-1}).
\end{equation*}
By Corollary 3.2 in \cite{bosq-00}, we have that
\begin{equation*}
  \norm{\frac{1}{T}\sum_{t=1}^Tw_{t-1}} = O(T^{-1/2}\ln^{1/2}T) \quad \text{a.s.}.
\end{equation*}
By Lemma \ref{lemma:1}, we have that $\norm{\bar f - \mathbb{E} f} = O(T^{-1/2}\ln^{1/2} T)$ a.s.. Then
\begin{equation*}
  \norm{\frac{1}{T}\sum_{t=1}^T (\hat w_t - w_t)\otimes w_{t-1}} = O_p(T^{-1}\ln T).
\end{equation*}

For the second term on the right hand side of \eqref{eq:R_2}, notice that by \eqref{eq:dw_t_MSE}, we have
\begin{align*}
  \mathbb{E}\norm{\frac{1}{T}\sum_{t=1}^T (\hat w_t - w_t) \otimes (\hat w_{t-1} - w_{t-1})} & \leq \frac{1}{T}\sum_{t=1}^T\left(\mathbb{E}\norm{\hat w_t-w_t}^2\right)^{1/2}\left(\mathbb{E}\norm{\hat w_{t-1}-w_{t-1}}^2\right)^{1/2} \\
  & = O(T^{-1}).
\end{align*}
This implies that
\begin{equation*}
  \norm{\frac{1}{T}\sum_{t=1}^T (\hat w_t - w_t) \otimes (\hat w_{t-1} - w_{t-1})} = O_p(T^{-1}).
\end{equation*}
Then we have that $\norm{R_2} = O_p(T^{-1}\ln T)$.

Similarly, one can show that $\norm{R_3} = O_p(T^{-1}\ln T)$. Then $\norm{R_1+R_2+R_3} = O_p(T^{-1}\ln T)$. Together with \eqref{eq:lambda_K} and that $\norm{w_T}\leq 2M$ a.s., this implies that
\begin{equation*}
  \sqrt{T/K}\norm{(R_1+R_2+R_3)\hat Q_K^+(w_T)} \leq \sqrt{T/K}\norm{R_1+R_2+R_3}\lambda_K^{-1}\norm{w_T} \to_p 0.
\end{equation*}
Then \eqref{eq:AN_1} holds.

To prove that \eqref{eq:AN_2} holds, note that
\begin{equation*}
  \norm{\hat Q_K^+ - Q_K^+} \leq \norm{E_2} + \norm{E_3}
\end{equation*}
where
\begin{equation*}
  E_2 = \sum_{k=1}^K \hat \lambda_k^{-1}(\hat v_k \otimes \hat v_k) - \sum_{k=1}^K \lambda_k^{-1}(\hat v_k \otimes \hat v_k),
\end{equation*}
and
\begin{equation*}
  E_3 = \sum_{k=1}^K \lambda_k^{-1}(\hat v_k \otimes \hat v_k) - \sum_{k=1}^K \lambda_k^{-1}(v_k \otimes v_k).
\end{equation*}
That $\norm{E_2}\to_p 0$ and $\norm{E_3}\to_p 0$ follow immediately from the proofs of $\norm{D_2}\to_{a.s.} 0$ and $\norm{D_3}\to_{a.s.} 0$ in Theorem \ref{thm:5}. Therefore, $\norm{\hat Q_K^+ - Q_K^+}\to_p 0$.

It follows from the proof of Lemma 4.6 in \citet{bosq-00} that $\norm{S} = O_p(1/\sqrt{T})$. Then
\begin{equation*}
   \sqrt{T/K}\norm{S(\hat Q_K^+ - Q_K^+)(w_T)} \leq \sqrt{T/K}\norm{S}\norm{\hat Q_K^+ - Q_K^+}\norm{w_T} \to_p 0.
\end{equation*}
That is, \eqref{eq:AN_2} holds.

By \eqref{eq:dw_t_MSE}, $\norm{\hat w_T - w_T} = O_p(1/\sqrt{T})$. Then \eqref{eq:AN_3} follows immediately. This completes the proof of the Lemma.
\end{proof}


\begin{proof}[Proof of Theorem \ref{thm:8}]
  Due to results in Lemma \ref{lemma:7}, it suffices to show that
  \begin{equation*}
    \sqrt{T/K}(A\hat \Pi_K w_T - Aw_T) = o_p(1).
  \end{equation*}
Since
\begin{align*}
  \norm{A\hat \Pi_K w_T - Aw_T} & \leq \norm{A\hat \Pi_K w_T - A\Pi_K w_T} + \norm{A\Pi_K w_T - Aw_T} \\
  & \leq \norm{A}\norm{\hat \Pi_K - \Pi_K}\norm{w_T} + \norm{A}\norm{(I-\Pi_K)w_T},
\end{align*}
it suffices to show that
\begin{equation}
  \label{eq:AN_4}
  \norm{\hat \Pi_K - \Pi_K} = o_p\left(T^{-1/2}K^{1/2}\right)
\end{equation}
and that
\begin{equation}
  \label{eq:AN_5}
  \norm{(I-\Pi_K)w_T} = o_p\left(T^{-1/2}K^{1/2}\right).
\end{equation}

To prove \eqref{eq:AN_4}, we need some tools from functional calculus. Let
\begin{equation*}
  \delta_k = \min\{\lambda_k - \lambda_{k+1}, \lambda_{k-1}-\lambda_{k}\}
\end{equation*}
and
\begin{equation*}
  \hat \delta_k = \min\{\hat \lambda_k - \hat \lambda_{k+1}, \hat \lambda_{k-1}-\hat \lambda_{k}\}
\end{equation*}
where we set $\lambda_0 = \hat \lambda_0 = +\infty$. Notice that by convexity of $\lambda_k$, we have that $\delta_k = \lambda_{k}-\lambda_{k+1}$. Let $B_k$ be the oriented circle in the complex plane with center $\lambda_k$ and radius $\delta_k/3$ and let $\hat B_k$ be the oriented circle in the complex plane with center $\hat \lambda_k$ and radius $\hat \delta_k/3$. Notice that the $B_k$'s are disjoint. Also, the $\hat B_k$'s are disjoint. Let $B = \bigcup_{k=1}^K B_k$ and $\hat B = \bigcup_{k=1}^K \hat B_k$. Then we have that
\begin{equation*}
  \Pi_K = \frac{1}{2\pi i}\oint_B (zI - Q)^{-1} dz
\end{equation*}
and
\begin{equation}
  \label{eq:hat_Pi_K}
  \hat \Pi_K = \frac{1}{2\pi i}\oint_{\hat B} (zI - \hat Q)^{-1} dz.
\end{equation}
The contour integrals here are defined as Stieltjes integrals on the complex plane whose convergences are in the norm of $\mathcal{L}(H)$, the space of all bounded linear operators on $H$. For details of functional calculus, one may see, for example, \citet[Chapter 1]{gohberg-goldberg-kaashoek-90}.

Next, we show that
\begin{equation}
  \label{eq:hatB=B}
  \frac{\sqrt{T}}{2\pi i}\oint_{\hat B} (zI - \hat Q)^{-1} dz \to_p \frac{\sqrt{T}}{2\pi i}\oint_{B} (zI - \hat Q)^{-1} dz.
\end{equation}
To show \eqref{eq:hatB=B}, we define
\begin{equation*}
  \mathcal{A}_T = \bigcap_{k=1}^K\left\{\omega\in \Omega: \abs{\hat \lambda_k(\omega) - \lambda_k} < \frac{\delta_k}{3} \right\}.
\end{equation*}
Note that for all $\omega\in \mathcal{A}_T$, $\hat \lambda_k(\omega)$ lies in the interior of the circle $B_k$ for all $k\in \{1, 2, \ldots, K\}$. This implies that we may replace all $\hat B_k$'s with $B_k$'s and therefore replace $\hat B$ with $B$ in \eqref{eq:hat_Pi_K} without affecting the value of the right hand side of \eqref{eq:hat_Pi_K} for all $\omega\in \mathcal{A}_T$. Then
\begin{align*}
  \frac{\sqrt{T}}{2\pi i}\oint_{\hat B} (zI - \hat Q)^{-1} dz =
  &  \frac{\sqrt{T}}{2\pi i}\oint_{B} (zI - \hat Q)^{-1} dz \\
  & + \left(\frac{\sqrt{T}}{2\pi i}\oint_{\hat B} (zI - \hat Q)^{-1} dz -  \frac{\sqrt{T}}{2\pi i}\oint_{B} (zI - \hat Q)^{-1} dz\right) 1\left(\mathcal{A}_T^C\right).
\end{align*}

Denote
\begin{equation*}
  L_1 = \left(\frac{\sqrt{T}}{2\pi i}\oint_{\hat B} (zI - \hat Q)^{-1} dz -  \frac{\sqrt{T}}{2\pi i}\oint_{B} (zI - \hat Q)^{-1} dz\right) 1\left(\mathcal{A}_T^C\right).
\end{equation*}
By Theorem 4.10 in \cite{bosq-00}, the asymptotic behavior of $\abs{\hat \lambda_k - \lambda_k}$ in distribution is the same as that of $\langle (\hat Q - Q)v_k, v_k\rangle$. Then for any $\eta>0$, we have that
\begin{align*}
  \mathbb{P}(\norm{L_1}>\eta) \leq \mathbb{P}\left(\mathcal{A}_T^C\right)
  & \leq \sum_{k=1}^K \mathbb{P}\left(\abs{\hat \lambda_k - \lambda_k}\geq \frac{\delta_k}{3}\right)\\
  & \leq \frac{3\lambda_k}{\delta_k}\mathbb{E}\frac{\abs{\hat \lambda_k -\lambda_k}}{\lambda_k} \\
  & \sim \frac{3\lambda_k}{\delta_k}\mathbb{E}\frac{\langle(\hat Q- Q)v_k, v_k\rangle}{\lambda_k}.
\end{align*}
To prove that the above term converges to 0, we first prove that
\begin{equation}
  \label{eq:sup_m_p}
  \sup_{m, p} T\frac{\mathbb{E} \langle ( \hat Q- Q )v_p, v_m\rangle^2}{\lambda_p\lambda_m} \leq M
\end{equation}
where $M$ is some constant. Write
\begin{equation*}
  \hat Q - Q = \tilde Q - Q +R,
\end{equation*}
where $\tilde Q$ has been defined in Theorem \ref{thm:2} and $R$ includes all residual terms. Then
\begin{equation*}
  \langle ( \hat Q- Q )v_p, v_m\rangle^2 = \langle ( \tilde Q- Q )v_p, v_m\rangle^2 + \langle Rv_p, v_m\rangle^2 + 2\langle (\tilde Q - Q)v_p, v_m\rangle \langle Rv_p, v_m\rangle.
\end{equation*}
It has been established in \cite{mas-07} that
\begin{equation*}
  \sup_{m, p} T\frac{\mathbb{E} \langle ( \tilde Q- Q )v_p, v_m\rangle^2}{\lambda_p\lambda_m} \leq M.
\end{equation*}
Other terms are negligible. Here we examine $\langle Rv_p, v_m\rangle^2$. For example, a typical term of $R$ is $R_0 = T^{-1}\sum_{t=1}^T \Delta_t\otimes w_t$. We have
\begin{align*}
  \langle R_0v_p, v_m\rangle^2
  & = \frac{1}{T^2}\left(\sum_{t=1}^T\langle w_t, v_p\rangle \langle \Delta_t, v_m\rangle \right)^2 \\
  & = \frac{1}{T^2}\sum_{t=1}^T \langle w_t, v_p\rangle^2\langle \Delta_t, v_m\rangle^2 + \frac{2}{T^2}\sum_{1\leq t< s\leq T} \langle w_t, v_p\rangle\langle \Delta_t, v_m\rangle \langle w_s, v_p\rangle \langle \Delta_s, v_m\rangle.
\end{align*}
We have
\begin{align*}
  \mathbb{E}\frac{1}{T^2}\sum_{t=1}^T \langle w_t, v_p\rangle^2\langle \Delta_t, v_m\rangle^2
  & \leq \frac{1}{T^2}\sum_{t=1}^T\left(\mathbb E \langle w_t, v_p\rangle^4\right)^{1/2}\left(\mathbb E\langle \Delta_t, v_m\rangle^4\right)^{1/2} \\
  & \leq \frac{M\lambda_m\lambda_p}{T}\sup_m\frac{\mathbb{E}\langle \Delta_1, v_m\rangle^2}{\lambda_m} = \frac{M\lambda_p\lambda_m}{T}O(N^{-r}).
\end{align*}
Also, note that
\begin{align*}
  & \abs{\mathbb{E} \frac{2}{T^2}\sum_{1\leq t< s\leq T} \langle w_t, v_p\rangle\langle \Delta_t, v_m\rangle \langle w_s, v_p\rangle \langle \Delta_s, v_m\rangle} \\
  \leq & \frac{2}{T^2}\sum_{1\leq t< s\leq T} Mb^{\abs{t-s}}\left[\mathbb{E} \left(\langle w_t, v_p\rangle^2\langle w_s, v_p\rangle^2 \right) \mathbb{E} \left(\langle \Delta_t, v_m\rangle^2 \langle \Delta_s, v_m\rangle^2 \right) \right]^{1/2}\\
  \leq & \frac{2}{T^2}\sum_{1\leq t< s\leq T} Mb^{\abs{t-s}}\left[ \left(\mathbb{E}\langle w_t, v_p\rangle\right)^4\left(\mathbb{E}\langle w_s, v_p\rangle\right)^4  \left(\mathbb{E}\langle \Delta_t, v_m\rangle\right)^4 \left(\mathbb{E}\langle \Delta_s, v_m\rangle \right)^4 \right]^{4}\\
  \leq & \frac{\lambda_m\lambda_p}{T}\sum_{j=1}^{T-1}\left(1 - \frac{j}{T}\right)b^jO(N^{-r}) = \frac{\lambda_m\lambda_p}{T}O(N^{-r}),
\end{align*}
Similar arguments apply to other terms in $R$. Now we may conclude \eqref{eq:sup_m_p} and therefore that
\begin{equation*}
  \frac{3\lambda_k}{\delta_k}\mathbb{E}\frac{\langle(\hat Q- Q)v_k, v_k\rangle}{\lambda_k} \leq \frac{M}{\sqrt{T}}\sum_{k=1}^K\frac{3\lambda_k}{\delta_k} \leq \frac{3M}{\sqrt{T}}\sum_{k=1}^Kk\ln k\leq \frac{3M}{\sqrt{T}}K^2\ln K.
\end{equation*}
By assumption, the last term converges to 0. This implies that $\lim_{T\to \infty}\mathbb{P}(\norm{L_1}>\eta)= 0$. Then \eqref{eq:hatB=B} holds. It then implies that
\begin{align*}
  \sqrt{T/K}\left( \hat \Pi_K - \Pi_K\right)
  & \to_p \sqrt{\frac{T}{K}}\left(\frac{1}{2\pi i}\oint_{B} \left[ (zI - \hat Q)^{-1} - (zI-Q)^{-1} \right] dz\right) \\
  & = \sqrt{\frac{T}{K}}\left(\frac{1}{2\pi i} \oint_{B} \left[ (zI - Q)^{-1}(\hat Q - Q) (zI-\hat Q)^{-1} \right] dz\right).
\end{align*}

Now consider
\begin{equation*}
  L_2 = (zI - Q)^{-1}(\hat Q - Q) (zI-\hat Q)^{-1} - (zI - Q)^{-1}(\hat Q - Q) (zI-Q)^{-1}
\end{equation*}
and write
\begin{equation}
\label{eq:L2_decomp}
\begin{aligned}
  L_2
  = & (zI - Q)^{-1}(\hat Q - Q) [(zI-\hat Q)^{-1}-(zI-Q)^{-1}] \\
  =  & (zI - Q)^{-1}(\hat Q - Q)(zI - Q)^{-1}(\hat Q - Q)(zI - \hat Q)^{-1} \\
  = & (zI - Q)^{-\frac{1}{2}}\left[ (zI - Q)^{-\frac{1}{2}}(\hat Q - Q) (zI - Q)^{-\frac{1}{2}}\right]^2 \\
  & \cdot \left[ (zI - Q)^{\frac{1}{2}} (zI - \hat Q) (zI - Q)^{\frac{1}{2}} \right](zI - Q)^{-\frac{1}{2}}.
\end{aligned}
\end{equation}
With minor modifications of proofs in \cite{mas-07}, we may show that for
\begin{equation*}
  \mathcal{E}_k = \left\{\sup_{z\in B_k} \norm{(zI - Q)^{-\frac{1}{2}}(\hat Q - Q) (zI - Q)^{-\frac{1}{2}}} < \frac{1}{2}\right\},
\end{equation*}
we have that
\begin{equation}
  \label{eq:bound_part_1}
  \sup_{z\in B_k}\norm{(zI - Q)^{\frac{1}{2}} (zI - \hat Q) (zI - Q)^{\frac{1}{2}}} 1(\mathcal{E}_k) \leq 2 \quad \text{ a.s. }
\end{equation}
and that
\begin{equation}
  \label{eq:bound_part_prob}
  \mathbb{P}(\mathcal{E}_k^C) \leq M\frac{k\ln k}{\sqrt{T}}.
\end{equation}
Also, for $k$ large enough, we have that
\begin{equation}
  \label{eq:bound_part_2}
  \mathbb{E} \sup_{z\in B_k} \norm{(zI - Q)^{-\frac{1}{2}}(\hat Q - Q) (zI - Q)^{-\frac{1}{2}}}^2 \leq \frac{M}{T}k^2 \ln^2 k.
\end{equation}
Equation \eqref{eq:bound_part_prob} implies that for any $\eta>0$,
\begin{equation}
\label{eq:L2_part1}
  \mathbb{P}\left( \sqrt{\frac{T}{K}}\frac{1}{2\pi i}\oint_{B} L_2 1\left(\bigcup_{k=1}^K \mathcal{E}_k^C\right)dz >\eta\right) \leq \sum_{k=1}^K \mathbb{P}(\mathcal{E}_k^C) \leq \frac{M}{\sqrt{T}}K^2 \ln K \to 0
\end{equation}
under the assumption. On the other hand, \eqref{eq:L2_decomp}, \eqref{eq:bound_part_1},  \eqref{eq:bound_part_2}, and the fact that $\norm{zI - Q}^{-1/2} \leq \delta_k^{-1/2}$ for all $k\in\{1, 2, \ldots K\}$ implies that
\begin{align*}
  \sqrt{\frac{T}{K}}\frac{1}{2\pi i}\oint_{B} L_2 1\left(\bigcap_{k=1}^K \mathcal{E}_k\right) dz
  & = \sqrt{\frac{T}{K}}\frac{1}{2\pi i}\sum_{k=1}^K \oint_{B_k} L_2 1\left(\mathcal{E}_k\right)dz \\
  & \leq \sqrt{\frac{T}{K}}\frac{1}{2\pi i}\sum_{k=1}^K \oint_{B_k} \delta_k^{-1/2}\frac{M}{T}k^2\ln^2 k \cdot 2\delta_k^{-1/2} dz \quad \text{ a.s. }\\
  & \leq \frac{M'}{\sqrt{T}}K^{5/2}\ln^2K \quad \text{ a.s.. }
\end{align*}
By assumption, the last term converges to 0. This, together with \eqref{eq:L2_part1} implies that
\begin{equation*}
  \sqrt{\frac{T}{K}}\frac{1}{2\pi i}\oint_{B} L_2 dz \to_p 0,
\end{equation*}
that is,
\begin{equation*}
  \sqrt{T/K}\left( \hat \Pi_K - \Pi_K\right) \to_p
  \sqrt{\frac{T}{K}}\left(\frac{1}{2\pi i} \oint_{B} \left[ (zI - Q)^{-1}(\hat Q - Q) (zI- Q)^{-1} \right] dz\right).
\end{equation*}

It is shown in \cite{dauxois-pousse-romain-82} that
\begin{equation*}
  \frac{1}{2\pi i}\oint_{B_k} (zI-Q)^{-1}(\hat Q - Q)(zI - Q)^{-1} dz = \psi_k (\hat Q - Q) \pi_k + \pi_k (\hat Q - Q) \psi_k
\end{equation*}
where $\pi_k = v_k\otimes v_k$ is the projection on the eigenspace associated with $\lambda_k$, and $\psi_k = \sum_{\ell \not =  k}\pi_\ell/(\lambda_\ell - \lambda_k)$. As a consequence, we have that
\begin{equation*}
  \langle v_i, (\hat \Pi_K - \Pi_K)v_j\rangle =
  \begin{cases}
    \displaystyle\frac{\langle v_i, (\hat Q - Q)v_j\rangle}{\lambda_i - \lambda_j}, & 1\leq i\leq K <j, \\
    \displaystyle\frac{\langle v_i, (\hat Q - Q)v_j\rangle}{\lambda_j - \lambda_i}, & 1\leq j\leq K <i,
  \end{cases}
\end{equation*}
with all other terms vanishing.

Then
\begin{align*}
  \norm{\hat \Pi_K - \Pi_K} & = \sup_{\norm{v}=1} \langle v, (\hat \Pi_K - \Pi_K)v\rangle \\
  & = \sup_{\norm{v} = 1} \sum_{i=1}^\infty \sum_{j=1}^\infty \langle v, v_i \rangle \langle v, v_j\rangle \langle v_i, (\hat \Pi_K - \Pi_K)v_j\rangle \\
  & = 2 \sup_{\norm{v} = 1}\sum_{i=1}^K\sum_{j=K+1}^\infty \langle v, v_i \rangle \langle v, v_j\rangle \frac{\langle v_i, (\hat Q - Q)v_j\rangle}{\lambda_i - \lambda_j} \\
  & = 2 \sup_{\norm{v} = 1}\sum_{i=1}^K\sum_{j=K+1}^\infty \langle v, v_i \rangle \langle v, v_j\rangle \frac{\langle v_i, (\tilde Q - Q)v_j\rangle + \langle v_i, Rv_j\rangle}{\lambda_i - \lambda_j}.
\end{align*}
We write
\begin{align*}
  & \sup_{\norm{v} = 1}\sum_{i=1}^K\sum_{j=K+1}^\infty \langle v, v_i \rangle \langle v, v_j\rangle \frac{\langle v_i, (\tilde Q - Q)v_j\rangle}{\lambda_i - \lambda_j} \\
  = & \sup_{\norm{v} = 1}\sum_{i=1}^K\sum_{j=K+1}^\infty \langle v, v_i \rangle \langle v, v_j\rangle \frac{1}{\lambda_i - \lambda_j}\frac{1}{T}\sum_{t=1}^T\langle v_i, (w_t\otimes w_t)v_j\rangle \\
  = & \sup_{\norm{v} = 1}\sum_{i=1}^K\sum_{j=K+1}^\infty \langle v, v_i \rangle \langle v, v_j\rangle \frac{\sqrt{\lambda_i\lambda_j}}{\lambda_i - \lambda_j}\frac{1}{T}\sum_{t=1}^T \frac{\langle v_i, w_t\rangle}{\sqrt{\lambda_i}} \frac{\langle v_j, w_t\rangle}{\sqrt{\lambda_j}} \\
  = & \sup_{\norm{v} = 1}\sum_{i=1}^K\sum_{j=K+1}^\infty \langle v, v_i \rangle \langle v, v_j\rangle \frac{\sqrt{\lambda_i\lambda_j}}{\lambda_i - \lambda_j}\frac{1}{T}\sum_{t=1}^T m_{ijt},
\end{align*}
where we have set
\begin{equation*}
  m_{ijt} = m_{it}m_{jt} = \frac{\langle v_i, w_t\rangle}{\sqrt{\lambda_i}} \frac{\langle v_j, w_t\rangle}{\sqrt{\lambda_j}}.
\end{equation*}
For all $i$ and $j$ such that $i\not = j$, $(m_{ijt})_{t\geq 1}$ is a mean zero stationary process. Moreover,
\begin{equation*}
  \mathbb{E} m_{ijt}^2 \leq \left(\mathbb{E} m_{it}^4\right)^{1/2}\left(\mathbb{E} m_{jt}^4\right)^{1/2} \leq \left(\mathbb{E} \frac{\langle v_i, w_t\rangle^4}{\lambda_i^2}\right)^{1/2} \left(\mathbb{E} \frac{\langle v_j, w_t\rangle^4}{\lambda_j^2}\right)^{1/2} \leq M,
\end{equation*}
and the autocovariance function $\Gamma_{ij}$ of $(m_{ijt})$ is absolutely summable. As a result, we have that
\begin{equation*}
  T\mathbb{E}\left( \frac{1}{T}\sum_{t=1}^T m_{ijt}\right)^2 = \frac{1}{T}\mathbb{E}\sum_{t=1}^T\sum_{s=1}^T m_{ijt}m_{ijs} = \sum_{k=1}^T\left(1 - \frac{k}{T}\right)\Gamma_{ij}(k) < \infty
\end{equation*}
for all $i\not =j$. Therefore, we have that
\begin{equation*}
  \left( \frac{1}{T}\sum_{t=1}^T m_{ijt}\right)^2 = O_p(T^{-1})
\end{equation*}
uniformly for all $i = 1, \ldots, K$ and $j = K+1, K+2, \ldots$. Obviously, for any $\norm{v} =1$, we have that
\begin{equation*}
  \sum_{i=1}^K\sum_{j=K+1}^\infty \langle v, v_i \rangle^2 \langle v, v_j\rangle^2 \leq \sum_{i=1}^K\langle v, v_i \rangle^2 \sum_{j=K+1}^\infty  \langle v, v_j\rangle^2 \leq 1.
\end{equation*}
Therefore, we have that
\begin{equation}
\label{eq:thm_8_sup}
\begin{aligned}
  & \abs{\sup_{\norm{v} = 1}\sum_{i=1}^K\sum_{j=K+1}^\infty \langle v, v_i \rangle \langle v, v_j\rangle \frac{\sqrt{\lambda_i\lambda_j}}{\lambda_i - \lambda_j}\frac{1}{T}\sum_{t=1}^T m_{ijt}} \\
  \leq & \left(\sum_{i=1}^K\sum_{j=K+1}^\infty \langle v, v_i \rangle^2 \langle v, v_j\rangle^2\right)^{1/2} \left(\sum_{i=1}^K\sum_{j=K+1}^\infty \frac{\lambda_i\lambda_j}{(\lambda_i - \lambda_j)^2}\left(\frac{1}{T}\sum_{t=1}^T m_{ijt}\right)^2 \right)^{1/2} \\
  = & o_p(T^{-1/2}K^{1/2})
\end{aligned}
\end{equation}
by assumption. That is,
\begin{equation*}
  \sup_{\norm{v} = 1}\sum_{i=1}^K\sum_{j=K+1}^\infty \langle v, v_i \rangle \langle v, v_j\rangle \frac{\langle v_i, (\tilde Q - Q)v_j\rangle}{\lambda_i - \lambda_j} = o_p(T^{-1/2}K^{1/2}).
\end{equation*}

The term
\begin{equation*}
  \sup_{\norm{v} = 1}\sum_{i=1}^K\sum_{j=K+1}^\infty \langle v, v_i \rangle \langle v, v_j\rangle \frac{\langle v_i, R v_j\rangle}{\lambda_i - \lambda_j}
\end{equation*}
can be analyzed analogously. For instance, one component of this term can be written as
\begin{align*}
  & \sup_{\norm{v} = 1}\sum_{i=1}^K\sum_{j=K+1}^\infty \langle v, v_i \rangle \langle v, v_j\rangle \frac{1}{\lambda_i - \lambda_j}\frac{1}{T}\sum_{t=1}^T\langle v_i, (w_t\otimes \Delta_t)v_j\rangle \\
  = & \sup_{\norm{v}=1} \sum_{i=1}^K\sum_{j=K+1}^\infty \langle v, v_i \rangle \langle v, v_j\rangle \frac{\sqrt{\lambda_i\lambda_j}}{\lambda_i - \lambda_j}\frac{1}{T}\sum_{t=1}^T \frac{\langle v_i, w_t\rangle}{\sqrt{\lambda_i}} \frac{\langle v_j, \Delta_t\rangle}{\sqrt{\lambda_j}}.
\end{align*}
Since
\begin{align*}
  \mathbb{E} \abs{\frac{1}{T}\sum_{t=1}^T \frac{\langle v_i, w_t\rangle}{\sqrt{\lambda_i}} \frac{\langle v_j, \Delta_t\rangle}{\sqrt{\lambda_j}}}
  & \leq \mathbb{E}\left(\frac{1}{T}\sum_{t=1}^T \frac{\langle v_i, w_t\rangle^2}{\lambda_i}\right)^{1/2}\left(\frac{1}{T}\sum_{t=1}^T \frac{\langle v_j, \Delta_t\rangle^2}{\lambda_j}\right)^{1/2}
 \\
  & \leq \left(\frac{1}{T}\sum_{t=1}^T \frac{\mathbb{E} \langle v_i, w_t\rangle^2}{\lambda_i}\right)^{1/2}\left(\frac{1}{T}\sum_{t=1}^T \frac{\mathbb{E} \langle v_j, \Delta_t\rangle^2}{\lambda_j}\right)^{1/2} \\
  & = O(N^{-r/2}),
\end{align*}
a similar argument as in \eqref{eq:thm_8_sup} shows that
\begin{equation*}
  \sup_{\norm{v}=1} \sum_{i=1}^K\sum_{j=K+1}^\infty \langle v, v_i \rangle \langle v, v_j\rangle \frac{1}{\lambda_i - \lambda_j}\frac{1}{T}\sum_{t=1}^T\langle v_i, (w_t\otimes \Delta_t)v_j\rangle = o_p(T^{-1/2}K^{1/2}).
\end{equation*}
Similarly, the other components could be shown to have order $o_p(T^{-1/2}K^{1/2})$. Therefore, we conclude that
\begin{equation*}
  \sup_{\norm{v} = 1}\sum_{i=1}^K\sum_{j=K+1}^\infty \langle v, v_i \rangle \langle v, v_j\rangle \frac{\langle v_i, R v_j\rangle}{\lambda_i - \lambda_j} = o_p(T^{-1/2}K^{1/2}),
\end{equation*}
and hence
\begin{equation*}
  \sqrt{T/K}\norm{\hat \Pi_K - \Pi_K} = o_p(1).
\end{equation*}

To complete the proof of the theorem, we only need to note that
\begin{align*}
  \mathbb{E}\norm{(I - \Pi_K)w_T}^2
  & = \mathbb{E}\norm{\left(\sum_{k=K+1}^\infty v_k\otimes v_k\right) w_T}^2 \\
  & = \mathbb{E}\norm{\sum_{k=K+1}^\infty \langle v_k, w_T\rangle v_k}^2 \\
  & = \mathbb{E}\left(\sum_{k = K+1}^\infty \langle v_k, w_T\rangle^2\right) \\
  & = \sum_{k=K+1}^\infty \lambda_k = o(T^{-1}K).
\end{align*}
\end{proof}


\newpage
\singlespacing
\bibliographystyle{elsevier}
\bibliography{far}

\end{document}